\documentclass[12pt,preprint]{aastex}  

\usepackage{a4}

\usepackage{graphicx}
\usepackage{lscape}

\newcommand{\sub}[1]{\ensuremath{_\textrm{\scriptsize{#1}}}}

\newcommand{\vgsr}{$V$\sub{GSR}}
\newcommand{\meanvgsr}{$\langle$$V$\sub{GSR}$\rangle$}
\newcommand{\kms}{km s$^{-1}$}
\newcommand{\degree}{\hbox{$^\circ$}}
\newcommand{\feh}{[Fe/H]}
\newcommand{\meanfeh}{$\langle$[Fe/H]$\rangle$}

\newcommand{\Lsun}{$\Lambda_{\odot}$}

\newcommand{\Bsun}{$B_{\odot}$}

\slugcomment{\emph{Accepted for Publication in the Astrophysical Journal}}

\shorttitle{EXPLORING THE SGR STREAM WITH SEKBO RRLS}
\shortauthors{PRIOR ET AL.}

\begin{document}

\title{Exploring the Sagittarius Stream\\
    with SEKBO Survey RR Lyrae Stars}

\author{Sayuri L. Prior, G. S. Da~Costa, Stefan C. Keller}
\affil{Research School of Astronomy and Astrophysics, Australian
  National University, Cotter Road, Weston Creek, Canberra, ACT 2611, Australia}

\begin{abstract}


A sample of RR Lyrae (RRL) variables from the Southern
Edgeworth-Kuiper Belt Object survey in regions overlapping the
expected position of debris from the interaction of the Sagittarius
(Sgr) dwarf galaxy with the Milky Way (RA $\sim$ 20 and 21.5 h;
distance = 16--21 kpc) has been followed up spectroscopically and
photometrically.  The 21 photometrically confirmed type $ab$ RRLs in
this region have \meanfeh\ = $-1.79 \pm 0.08$ on our system,
consistent with the abundances found for RRLs in a different portion
of the Sgr tidal debris stream.  The distribution of velocities in the
Galactic standard of rest frame (\vgsr) of the 26 RRLs in the region
is not consistent with a smooth halo population.  Upon comparison with
the Sgr disruption models of \citet{LJM05}, a prominent group of five stars 
having
highly negative radial velocities (\vgsr\ $\sim -175$ \kms) is
consistent with predictions for old trailing debris when the Galactic
halo potential is modeled as oblate.  In contrast, the prolate model 
does not predict any significant number of Sgr stars at the locations of
the observed sample.  The observations also require that the recent
trailing debris stream has a broader spread perpendicular to the Sgr
plane than predicted by the models.  We have also investigated the
possible association of the Virgo Stellar Stream (VSS) with Sgr debris by
comparing radial velocities for RRLs in the region with the same models, 
finding similarities in the velocity-position trends.  As suggested by
our earlier work, the stars in the VSS region with large negative \vgsr\ 
values are likely to be old leading Sgr debris, but we find that while 
old trailing Sgr debris may well make a contribution at positive \vgsr\ 
values, it is unlikely to fully account for the VSS feature.  Overall we 
find that further modeling is needed, as trailing arm data generally favors 
oblate models while leading arm data favors prolate models, with no single 
potential fitting all the observed data.

\end{abstract}
    


\section{Introduction}  \label{intro_sgr}

The notion that the process of galaxy formation involves a protracted,
dissipationless merging of protogalactic fragments, consistent with
the proposal of \citet{SZ78}, is now widely accepted and is consistent
with currently favored $\Lambda$~cold dark matter ($\Lambda$CDM)
cosmologies.  These cosmologies propose that galaxies form via the
hierarchical assembly of subgalactic dark halos and the subsequent
accretion of cooled baryonic gas \citep[e.g.][ and references
therein]{SFW06}.  The outer halo of our galaxy presents an excellent
opportunity for probing its formation due to its remoteness and
relative quiescence, and has consequently been frequently targeted in
searches for relic substructure arising from this accretion process.
\citet{BZB08} gauged quantitatively the relative importance of the
accretion mechanism in halo formation by comparing the level of
substructure present in Sloan Digital Sky Survey (SDSS) data to
simulations, and found that the data are consistent with a halo
constructed entirely from disrupted satellite remnants.  \citet{SH09}
reached a similar conclusion from their pencil-beam spectroscopic survey.
Systems undergoing disruption in the halo of the Gaxy have indeed been 
observed, with the most
striking example being the Sagittarius (Sgr) dwarf galaxy, located a
mere 16 kpc from the Galactic center and showing, through its
elongated morphology, unmistakable signs of a strong interaction with
the Galaxy \citep{IGI94}.  The debris from the interaction has
subsequently been observed around the sky
\citep{MSW03,NYG03,BZE06}, making it arguably the most significant
known contributor to the Galactic halo.


Since the discovery of the Sgr dwarf, many studies have reported
detections of suspected Sgr tidal debris streams using various tracers
including carbon stars from the APM survey \citep{TI98,ILI01}, red
clump stars from a pencil beam survey \citep{MSK99}, RR Lyrae stars
from the SDSS \citep{IGF00, WEB09} and from the Quasar Equatorial Survey Team
\citep[QUEST;][]{VZG05}, giant stars from the Spaghetti Project Survey
\citep{DHM01,SH09}, A type stars from the SDSS \citep{NYG03} and M giants
from the Two Micron All Sky Survey \citep[2MASS;][]{MSW03}.  Debris
associated with Sgr has been found at various angles from the current
position of the dwarf, as close as a few kpc from the sun
\citep{KMR02} and as far as the distant, outer reaches of the halo
\citep{NYG03}.  Some of the detections were hypothesized to be from
older wraps, lost on pericentric passages several Gyr ago
\citep{DHM01,KMR02,SH09}.

Stars stripped from the Sgr dwarf on recent orbits can be
unambiguously identified as part of the debris stream via their
highly constrained positions and velocities, but the association of
older debris can be more difficult and prone to debate.  The
overdensity in Virgo
\citep{VZA01,VZA04,VZ06,NYR02,DZV06,NYC07,JIB08,VJZ08,KMP08,PDK08},
hereafter referred to by \citeauthor{DZV06}'s nomenclature as the
Virgo Stellar Stream (VSS), is one such example.  The VSS has been
widely assumed to be a halo substructure which is independent of the
Sgr debris, though their association has been hypothesized by
\citet{MPJ07} who showed that \citeauthor{LJM05}'s \citeyearpar{LJM05}
model of the Sgr leading tidal tail passes through the region of the
VSS.  However, the model predicts highly negative radial velocities in
the Galactic standard of rest (GSR) frame for Sgr stars in this
region, contrary to observations of a peak velocity at $\vgsr \sim
100$--$130$ \kms\ \citep{DZV06,NYC07,PDK08}.  A possibility which has
received little attention thus far is the association of the VSS with
the \textit{trailing} Sgr debris stream, whose members are indeed
predicted to have positive velocities in this region.
\citeauthor{MPJ07}, commenting that in certain models both leading and
trailing arms overlap the VSS region, noted this possibility only in
passing.  On the other hand, the models predict a relatively low
density of Sgr debris in this region which is at odds with the
significance of the observed overdensity. \citeauthor{NYC07}
\citeyearpar{NYC07} also note that the VSS is not spatially coincident
with the main part of the Sgr leading tidal tail, but that the
features do significantly overlap.  It is thus clear that no consensus
has yet been reached regarding the association of the VSS with the Sgr
Stream.   
The current study addresses this question in \S\ref{rv_sgr} with intriguing 
results \citep[see also][]{SH09}.


As alluded to above, the wealth of observational data for potential
Sgr Stream members has motivated several attempts at modeling the
disruption of the Sgr dwarf and predicting the positions and radial
velocities of the ensuing debris particles
\citep[e.g.][]{JMS99,ILI01,HW01,MGA04,LJM05,FBE06}.  These models have
resulted in estimates of the orbit pericenter and apocenter, Sgr's
transverse velocity and current bound mass.  In addition, several
studies, considering the debris as test particles, have drawn
conclusions about the shape of the Galactic halo potential.  For example, 
from the observation that the Sgr stream
identified in their carbon star sample traces a great circle,
\citet{ILI01} concluded that the halo must be almost spherical.
Moreover, on the basis of N-body simulations, they subsequently
constrained the mass distribution in the dark halo and concluded that
it too was most likely spherical.  However,
\citet{Helmi04} asserted that debris lost recently, such as M giant
data in the trailing arm, is too young dynamically to provide any
constraints on the shape of the dark halo potential.  Nevertheless debris 
lost at earlier times can discriminate between prolate
(flattening $q > 1$), spherical ($q=1$) and oblate ($q<1$) shapes for
the potential \citep{Helmi04}.  

The issue of dark halo shape remains clouded, however, since a range
of contrasting views have also been presented in recent years.
\citet{Helmi04b} found that results from M giants in the leading arm
supported a notably prolate halo ($q=1.7$).  On the other hand,
\citet{MGA04}, using the various reported distances and radial
velocities available in the literature, obtained an oblate halo
($q=0.85$) in their simulations.  Based on RR Lyrae stars located in
the leading arm at a distance of $\sim$50 kpc, \citet{VZG05} found that
the models of \citet{MGA04} and \citet{Helmi04} with spherical or
prolate halos fit the data better than those for an oblate halo.
\citet{LJM05} performed the first simulations based on the 2MASS
all-sky view of the Sgr debris streams.  They noted that the
velocities of leading debris were better fit by prolate halos, whereas
trailing debris velocity data had a slight preference for oblate
halos.  Confusing the matter, however, was the observation that the
orbital pole precession of young debris favors oblate models.
\citet{FBE06} found that only halos close to spherical
resulted in bifurcated streams, such as that evident in
\citeauthor{BZE06}'s \citeyearpar{BZE06} analysis of upper main
sequence and turn off stars in SDSS data.  Finally \citet{SH09} favor
spherical or prolate halo shapes when comparing the Sgr debris in
their pencil-beam survey with the models of \citet{Helmi04}.

The brief discussion above illustrates that the shape of the dark halo
is a point of contention, with no model being capable of fitting all
the available data.  \citet{LJM05} propose that evolution of the
orbital parameters of Sgr over several Gyr is needed to explain the
results and suggest dynamical friction as the most likely mechanism to
bring this about.  They note that this would require Sgr to have been
an order of magnitude more massive 2 Gyr ago.  They also recognize the
need to model the dwarf as a two-component system, with the dark
matter bound less tightly than the baryons and indicate that such a
study is underway.  Inclusion of additional, newly discovered samples
of observed Sgr stream stars will no doubt help to more tightly
constrain these improved models, particularly samples containing old
stars which potentially were lost several pericentric passages ago and
which have been influenced by the Galactic halo potential for a longer
time period.

Being Population II stars, RR Lyrae (RRL) variable stars are an
excellent choice of tracer for unraveling the history of the Sgr-Milky
Way interaction and for further elucidating the shape of the dark
halo.  We have undertaken a project using RRLs which encompasses these
aims, while more generally seeking to gain further insight into the
role of accretion in galaxy formation.  Although our broad goal was to
identify and characterize substructures in the Galaxy's halo,
ultimately the project transpired to focus on two such substructures
in detail: the Virgo Stellar Stream and the Sagittarius Stream.  A
companion paper \citep[][ hereafter Paper I]{PDK08} describes in detail
the entire set of observations, along with results and analysis for
the VSS region.  In the current paper, results for our RRL samples in
the Sgr region are presented.  Specifically, after giving a brief
overview of the observations in \S\ref{obs}, radial velocities for
RRLs in the Sgr region are presented in \S\ref{rv_sgr}, and
comparisons are made to \citeauthor{LJM05}'s \citeyearpar{LJM05}
recent N-body simulations of the disruption of the Sgr dwarf, which
adopt various modeled shapes for the dark halo potential.  At this
point, data from the VSS region are also included in order to explore
their possible association with Sgr debris.  \S\ref{metal_sgr} then
presents metallicities of the Sgr region RRLs.  The paper ends with a
general discussion and conclusions.

\section{Observations} \label{obs}


The reader is referred to Paper I for a detailed account of target
selection, observations and data reduction.  Key points are reiterated
below.  The radial distribution of RRL candidates obtained by
\citet[][hereafter KMP08]{KMP08} from the Southern Edgeworth-Kuiper
Belt Object (SEKBO) survey data revealed several overdensities
compared to simulated smooth halos.  The two main overdensities
selected for follow-up spatially coincided with
the VSS (observed sample: 8 RRLs at RA $\sim$ 12.4 h and 3 RRLs at RA
$\sim$ 14 h; Paper~I), and a portion of the Sgr debris stream (observed 
sample: 5 RRLs at RA $\sim$ 20 h and 21 RRLs at RA $\sim$ 21.5 h; this paper).
The stars in both samples have
extinction corrected magnitudes $V_0 \sim 17$.  Although the possible
association of the VSS with Sgr debris is investigated here, for
ease of reference and for consistency with Paper I, we
refer to the samples of RRLs at $\sim$12.4 h and $\sim$14 h as being
in the ``VSS region'' and to the samples at $\sim$20 h and $\sim$21.5 h 
as being in the ``Sgr region''.

\begin{figure}
\centering
\includegraphics[width=0.8\textwidth]{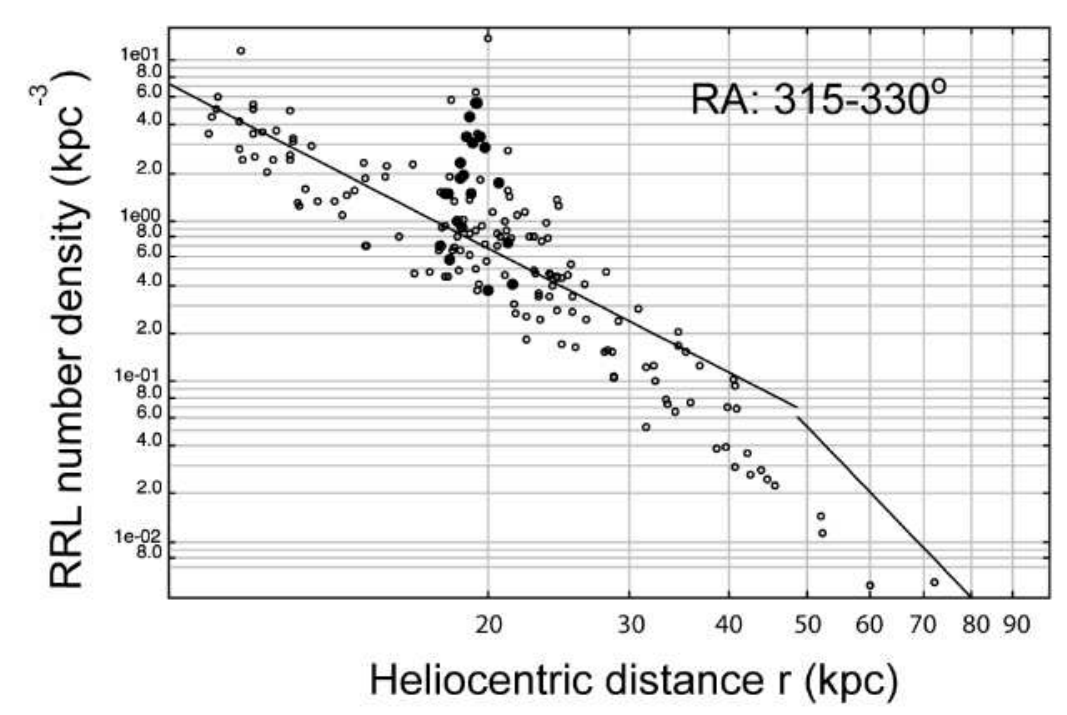}
\caption{The number density distribution of SEKBO RRL candidates for
the RA range 21 h -- 22 h as a function of heliocentric distance.  
The open circles are the individual data points and the filled circles
are the stars observed in the 21.5 h region.  The solid lines are the
smooth halo density distribution derived from the entire SEKBO sample
(see \citet{KMP08} for details). The marked excess
at $\sim$20 kpc is due to Sgr. \label{new_fig1} }
\end{figure}

In particular, we show in Fig.\ \ref{new_fig1} the number density of 
SEKBO RRL candidates as a function of heliocentric distance for the
Right Ascension range 21.0 h $\leq$ RA $\leq$ 22.0 h.  There is a clear
excess at distance of $\sim$20 kpc above the number density 
predicted for a smooth halo distribution, which KMP08 ascribe to Sgr debris.  
The ``Sgr region'' RRLs observed here
have a mean distance of 19.1 kpc with a range of 16 to 21 kpc, and clearly
probe this density excess.

As discussed in Paper I, photometric and spectroscopic observations
were made at Siding Spring Observatory in six six-night runs between
November 2006 and October 2007, with the Australian National
University $40''$ and 2.3 m telescopes, respectively.  For photometry,
5--19 (average 9) observations per star enabled confirmation of the RR
Lyrae classifications through period and light curve fitting in
addition to providing ephemerides.  For spectroscopy, 1--4
observations were taken for each target using the blue arm of the
Double Beam Spectrograph.  Spectra were centred on $4350$ \AA\ and
have a signal-to-noise of $\sim$20 and a resolution of $\sim$2 \AA.
The photometric and spectroscopic data reduction procedures are
described in Paper I and a photometric data summary, including
positions, magnitudes, periods and amplitudes of all targets, is
provided in Table 2 of that paper.

\section{Radial Velocities}  \label{rv_sgr}


The general method of obtaining radial velocities consisted of cross
correlating each wavelength-calibrated target spectrum with several
template spectra (radial velocity standard stars, for example).  For
each type $ab$ RRL, the velocities corresponding to different
epochs were then phased using ephemerides from the photometric data,
and a systemic velocity was subsequently assigned by fitting a
template RRL velocity curve (characterizing the RR$ab$ variation of
velocity with phase).  The average velocity was used as the systemic
velocity for type $c$ RRLs.  Finally, the heliocentric systemic
velocity was converted to a velocity in the Galactic standard of rest
frame.  The uncertainty in the \vgsr\ values is $\pm 20$ \kms.
Full details of the method are given in Paper I.

\begin{figure}[htbp]
~\centering
\includegraphics[width=0.95\textwidth]{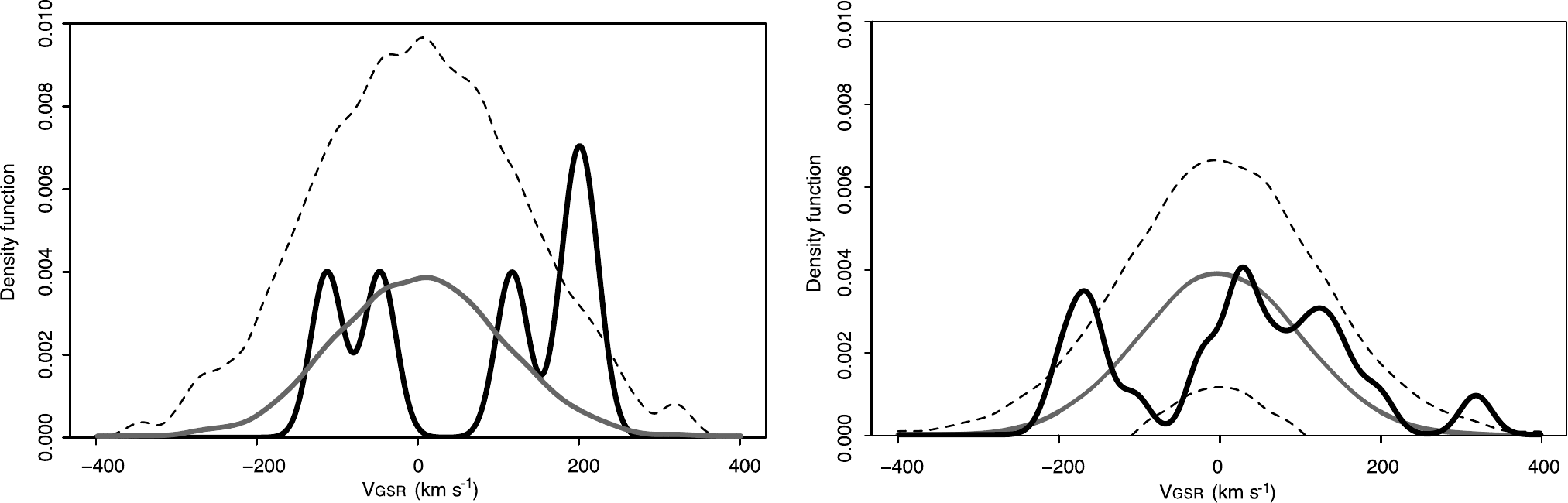}
        \caption{Generalized histograms of
          \vgsr\ (with kernel of 20 \kms) for observed RRLs in the Sgr
          region.  \textit{Left:} the 5 RRLs in the 20 h region;
          \textit{Right:} the 21 RRLs in the 21.5 h region.
          Overplotted in each panel as a solid gray line is the expected
          distribution of \vgsr\ for a halo population of the same size
          as the observed samples in which
          velocities are normally distributed with \meanvgsr\ = 0
          \kms\ and $\sigma$ = 100 \kms.  The 95\% confidence intervals
          for samples drawn from this distribution are shown by the dashed 
          lines.
         In the 20 h region, there is a statistically significant
          excess of stars with highly positive velocities (\vgsr\
          $\sim 200$ \kms) while in the 21.5 h region, there is a similar
          statistically significant excess of stars with highly negative 
          velocities (\vgsr\ $\sim -175$ \kms).}
        \label{genhisto_sgr}        
\end{figure}

Table \ref{spectrosgr} summarizes results from the spectroscopic data
for the 26 RRLs in the Sgr region, including the systemic \vgsr\
values\footnote{Positions, magnitudes, periods and amplitudes for the
stars are given in Table 2 of Paper I.}.
Results for the VSS region stars are also included for ease
of reference; see following discussion.  The distributions of \vgsr\
values in the 20 h (5 stars) and in the 21.5 h (21 stars) Sgr regions
are presented as generalized histograms in Fig.\ \ref{genhisto_sgr}
(note that results from the two regions are not combined since the
pattern of radial velocities of Sgr Stream stars varies with position,
as discussed further below).  Shown also in the figure are the expected
distributions for a halo RRL velocity distribution that has
\meanvgsr\ = 0 \kms\ and $\sigma$ = 100 \kms \citep[e.g.,][]{SGK04,BGK05}.
This ``smooth halo" velocity distribution is normalised with the assumption 
that the observed samples are entirely drawn from such a halo population  
though Fig.\ \ref{new_fig1} suggests this is not likely to be the case.
In the 20 h region, there is an apparent excess of stars
having highly positive radial velocities (\vgsr\ $\approx 200$ \kms),
while in the 21.5 h region an excess (five stars) is evident
at highly negative velocities (\vgsr\ $\approx -175$ \kms).  There is also 
an apparent lack of stars with $-125 <$ \vgsr\ $< -50$ \kms.

To investigate the statistical significance of these features we have
conducted Monte-Carlo simulations in which samples of size equal to the
observed samples were drawn randomly from the assumed smooth halo 
velocity distribution,
and convolved with a 20 \kms\ kernel as for the observations.  The results
are also displayed in Fig.\ \ref{genhisto_sgr} where the upper and lower 
contours that enclose 95\%  of the trials are shown by the dotted lines.  These
simulations show that in the 20~h region, the excess at \vgsr\ $\approx 
200$ \kms\ is indeed significant, as is the excess at \vgsr\ $\approx -175$ 
\kms\ in the 21.5~h region, even with the assumption that the
samples are entirely drawn from the smooth halo distribution.  In conjunction
with the  
results of Fig.\ \ref{new_fig1} in which there is a clear excess above the
smooth halo model, the results suggest strongly that both samples contain  
velocity components that are not simply random selections from a smooth halo.
The apparent deficiency of stars centered at \vgsr\ $\approx$ --75 \kms\ 
in the 21.5~h sample also appears to be statistically significant.

\begin{landscape}
\begin{deluxetable}{lrrcccrrcc}
\tablewidth{0pt}
\tabletypesize{\small}
\tablecaption{Spectroscopic data summary\label{spectrosgr}}
\tablehead{
\colhead{ID} &\colhead{\Lsun} &\colhead{\Bsun}
&\colhead{RRL}   &\colhead{systemic vel.}
&\colhead{n\sub{vel}}   &\colhead{$V$\sub{helio}}   &\colhead{\vgsr}
&\colhead{n\sub{Fe/H}}   &\colhead{[Fe/H]}
\\
& \colhead{(deg)} & \colhead{(deg)}
&\colhead{type}   &\colhead{calculation}
&\colhead{}   &\colhead{(\kms)}   &\colhead{(\kms)}
&\colhead{}   &\colhead{} 
}

\startdata

\multicolumn{10}{c}{\textsc{-----Sgr region-----}}\\
109582-3724  & $10.6$ & $-10.5$ & $ab$ & fit     & 3 & $129$  & $211$  & 4 & $-1.93$ \\
109464-5627  & $10.9$ & $-10.2$ & $ab$ & average & 4 & $36$   & $117$  & - & - \\
109489-3076  & $10.9$ & $-10.4$ & $ab$ & fit     & 3 & $-129$ & $-47$  & 3 & $-1.85$ \\
109489-2379  & $11.1$ & $-10.2$ & $ab$ & fit     & 3 & $108$  & $191$  & 3 & $-2.17$ \\ 
109463-3964  & $11.5$ & $-10.8$ & $ab$ & fit     & 2 & $-197$ & $-113$ & 3 & $-1.58$\\
125857-20    & $27.1$ & $-16.2$ & $ab$ & fit     & 1 & $-1$   & $109$  & 1 & $-2.00$ \\ 
110828-1100  & $29.3$ & $-10.7$ & $ab$ & fit     & 3 & $-250$ & $-159$ & 3 & $-1.76$ \\ 
126040-78    & $29.5$ & $-8.9$  & $ab$ & fit     & 2 & $-50$  & $34$   & 2 & $-1.45$ \\ 
110823-641   & $29.7$ & $-10.9$ & $c$  & average & 3 & $-112$ & $-20$  & - & - \\ 
114793-1530  & $30.2$ & $-16.1$ & $c$  & average & 3 & $51$   & $162$  & - & - \\ 
126093-1135  & $30.3$ & $-8.0$  & $ab$ & fit     & 3 & $56$   & $137$  & 3 & $-1.53$ \\ 
126536-714   & $31.0$ & $-7.2$  & $ab$ & fit     & 1 & $55$   & $134$  & 3 & $-0.73$ \\ 
99747-73     & $31.6$ & $-11.9$ & $ab$ & fit     & 1 & $-77$  & $20$   & 3 & $-2.55$ \\
102297-1488  & $32.3$ & $-13.6$ & $ab$ & fit     & 3 & $-290$ & $-187$ & 4 & $-1.53$ \\
110735-595   & $32.5$ & $-15.2$ & $ab$ & average & 4 & $-277$ & $-169$ & 4 & $-1.62$ \\
102292-1096  & $33.4$ & $-14.4$ & $ab$ & fit     & 2 & $92$   & $198$  & 2 & $-1.61$ \\
126245-763   & $33.7$ & $-8.7$  & $ab$ & fit     & 1 & $-13$  & $72$   & 1 & $-1.56$ \\ 
102601-1489  & $35.1$ & $-13.0$ & $ab$ & fit     & 2 & $-123$ & $-22$  & 3 & $-2.26$ \\
102601-400   & $35.3$ & $-12.8$ & $ab$ & fit     & 2 & $-301$ & $-201$ & 3 & $-1.98$ \\
110753-346   & $35.7$ & $-12.3$ & $ab$ & fit     & 2 & $221$  & $319$  & 3 & $-2.00$ \\
110827-579   & $36.5$ & $-11.5$ & $ab$ & fit     & 3 & $-252$ & $-157$ & 2 & $-1.51$ \\ 
99752-96     & $36.5$ & $-12.7$ & $c$  & average & 3 & $-74$  & $26$   & - & - \\
110738-411   & $37.2$ & $-12.0$ & $ab$ & fit     & 4 & $-206$ & $-107$ & 4 & $-2.10$ \\ 
113345-1032  & $37.6$ & $-11.5$ & $c$  & average & 3 & $7$    & $103$  & - & - \\ 
115381-767   & $38.8$ & $-12.5$ & $ab$ & fit     & 2 & $-74$  & $26$   & 3 & $-2.01$ \\
115381-349   & $39.4$ & $-12.6$ & $ab$ & fit     & 1 & $-38$  & $63$   & 2 & $-1.87$ \\

\multicolumn{10}{c}{\textsc{-----VSS region-----}}\\
108227-529  & $262.5$ & $17.8$  & $c$  & average & 1 & $-8$  &  $-119$ &  - &     - \\
96102-170   & $263.7$ & $15.2$  & $ab$ & fit     & 2 & $230$ &  $128$  &  3 &  $-2.15$ \\
120185-77   & $264.7$ & $13.0$  & $ab$ & fit     & 2 & $94$  &  $ 1$   &  2 &  $-2.38$ \\
107552-323  & $268.2$ & $18.0$  & $ab$ & fit     & 3 & $302$ &  $193$  &  3 &  $-1.34$ \\
119827-670  & $269.0$ & $17.7$  & $ab$ & fit     & 2 & $-84$ &  $-192$ &  2 &  $-1.68$ \\
120679-336  & $269.8$ & $19.5$  & $ab$ & fit     & 3 & $204$ &  $91$   &  2 &  $-1.74$ \\
121194-205  & $273.4$ & $20.0$  & $c$  & average & 1 & $-39$ &  $-152$ &  - &     - \\
121242-188  & $274.3$ & $22.4$  & $c$  & average & 3 & $276$ &  $155$  &  - &     - \\
120698-392  & $285.9$ & $16.8$  & $c$  & average & 1 & $227$ &  $134$  &  - &     - \\
109247-528  & $286.4$ & $14.7$  & $c$  & average & 2 & $113$ &  $28$   &  - &     - \\
105648-222  & $287.0$ & $11.2$  & $ab$ & fit     & 1 & $-91$ &  $-162$ &  1 &  $-1.45$ \\

\enddata


\end{deluxetable}
\end{landscape}

\begin{figure}[htbp]      
~\centering
\includegraphics[width=0.8\textwidth]{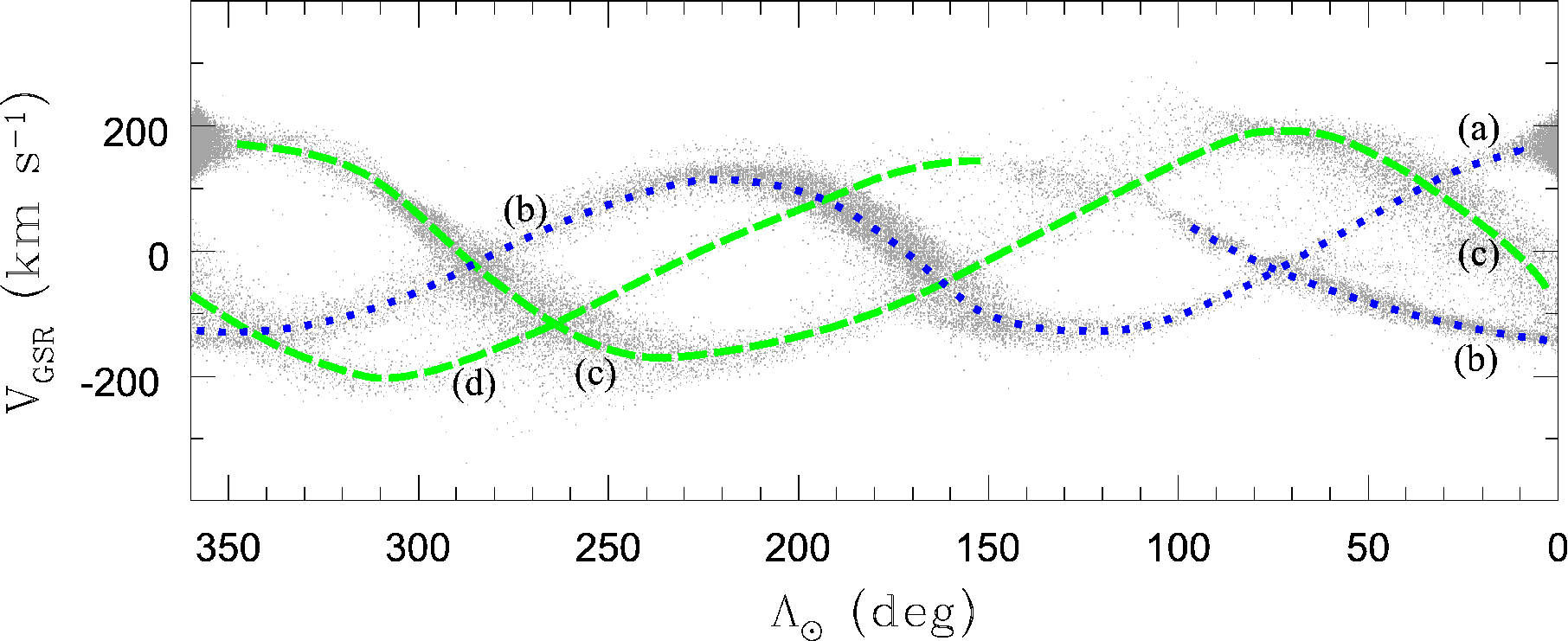}
        \caption[Radial velocities against \Lsun\
       from \citeauthor{LJM05}'s \citeyearpar{LJM05} model with a
       prolate Galactic halo potential.  Leading and trailing arms are
       highlighted and certain parts of the stream have been
       identified for ease of reference.]{Radial velocities against
       \Lsun\ from \citeauthor{LJM05}'s \citeyearpar{LJM05} model with
       a prolate Galactic halo potential.  Particles at all distances
       are shown (see Figs \ref{allmodels} and \ref{allmodelsd1621}
       for distance-coding). By definition, the Sgr dwarf is located
       at \Lsun\ = 0\degree.  The leading arm is highlighted by the
       green dashed line and the trailing arm by the blue dotted line.
       For ease of reference, certain parts of the stream have been
       identified: (a) new trailing debris (current perigalactic
       passage); (b) very old trailing debris (4 orbits ago); (c) old
       leading debris (3--4 orbits ago); and (d) very old leading
       debris (4 orbits ago). All \Bsun\ values are shown. }
        \label{streams}        
\end{figure}

In order to further investigate the pattern of radial velocities and
to explore how well they agree with expectations for Sgr Stream stars,
we have compared the velocities to those predicted by the recent
models of \citet{LJM05}\footnote{\citet{LJM05} provide their data at
http://www.astro.virginia.edu/$\sim$srm4n/Sgr.} (hereafter LJM05).  While a
number of other models for the disruption from the Sgr-Milky Way
interaction exist (see \S\ref{intro_sgr} for a brief review), LJM05's models
were chosen for the comparison as they are currently the only models
based on a complete all-sky view of the tidal streams of Sgr (namely,
the 2MASS M giant sample).  They use N-body simulations to predict the
heliocentric distances and radial velocities of $10^5$ particles
stripped from the Sgr dwarf up to four orbits ago (i.e.\ up to
approximately 3 Gyr ago).  Three models of the flattening, $q$, of the
Galactic dark halo potential are considered: prolate ($q$ = 1.25),
spherical ($q$ = 1.0) and oblate ($q$ = 0.90).  Their simulations
adopt a spherical, Sun-centered, Sgr coordinate
system\footnote{We converted from the standard Galactic coordinate
system to the Sgr longitudinal coordinate system using David
R. Law's C++ code, provided at the website above.}, with the 
longitudinal coordinate, \Lsun, being
zero in the direction of the Sgr core and increasing along the Sgr
trailing debris stream.  The zero plane of the latitude coordinate,
\Bsun, is defined by the best fit great circle of the Sgr debris. For
the purpose of clarifying the discussion which follows, Fig.\ 
\ref{streams} shows \vgsr\ plotted against \Lsun\ for the case of the
prolate potential (selected due to its more easily identifiable
streams compared to oblate and spherical potentials). All \Bsun\ values
are shown.  The Sgr core
is, by definition, located at \Lsun\ = 0\degree, the leading stream is
highlighted by the green dashed line and the trailing stream by the
blue dotted line.  In addition, certain parts of the streams have been
labeled (a--d) for future reference.  In the text which follows, they
are referred to as, for example, \textit{``debris-c''}, where the
\Lsun\ range should also be taken into account in order to focus on
the relevant stream section.

\begin{figure}[htbp]    
~\centering
\includegraphics[width=0.95\textwidth]{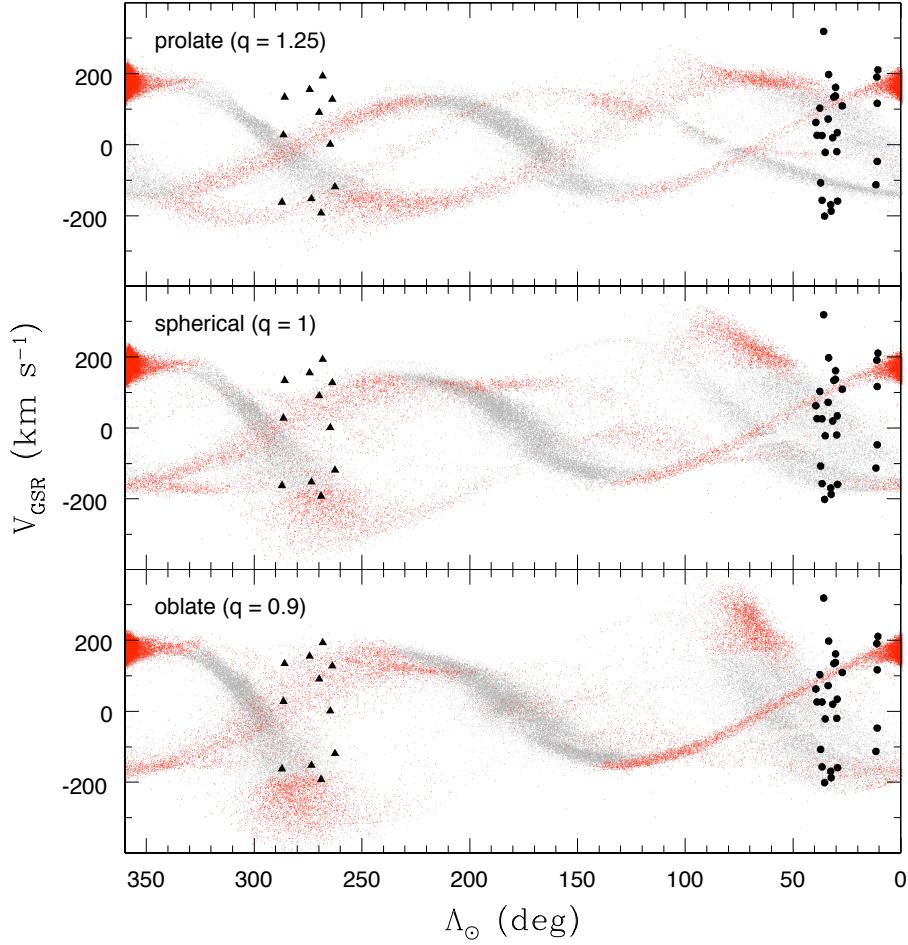}
        \caption[Radial velocities against $\Lambda_{\odot}$ from
          models with prolate, spherical and oblate Galactic halo
       potentials \citep{LJM05}, with values for hypothesized Sgr
       Stream members observed in the Sgr and VSS regions overplotted.
       Model distances 6--31 kpc are highlighted.]{Radial velocities
       against $\Lambda_{\odot}$ from models with prolate
       \textit{(top)}, spherical \textit{(middle)} and oblate
       \textit{(bottom)} Galactic halo potentials \citep{LJM05}.
       Red dots denote simulated
       stars with heliocentric distances $d$ = 6--31 kpc while gray
       denotes stars at all other distances.  Overplotted are values
       for the 26 observed RRLs in the 20 h and 21.5 h Sgr regions
       \textit{(circles)} and the 11 RRLs in the VSS region
       \textit{(triangles)}.  Note that, at the relevant distance,
       only the oblate model is able to predict the group of stars at
       \Lsun\ $\approx$ 35\degree\ which are observed to have highly
       negative \vgsr. All \Bsun\ values are shown. }
        \label{allmodels}        
\end{figure}

Figure \ref{allmodels} shows the distribution of \vgsr\ as a function
of \Lsun\ for LJM05's simulated particles assuming prolate, spherical
and oblate halo potentials.  All \Bsun\ values are shown. 
Our observed RRLs are at heliocentric
distances 16--21 kpc.  The distances are based on the assumption of $M_V$
= 0.56 and have an uncertainty of $\sim$7\%, as described in KMP08.
This corresponds to an uncertainty of approximately $\pm$1 kpc at a
distance of 20 kpc.  In order to isolate simulated stars at similar
distances to the observed stars, the plot is color-coded, with red
dots representing particles at 6--31 kpc
(i.e.\ 10 kpc closer/farther than our observed range).  All other
distances are color-coded gray.  The choice to consider a wider
distance range for the simulated particles than for our observational
data was motivated by the significant uncertainties involved in the
modeled distances.  The major factor contributing to this is the 17\%
distance uncertainty for 2MASS M giant stars, stemming from the
uncertainty in the values of the solar distance from the Galactic
center and from Sgr.  LJM05 note that the estimated size of Sgr's
orbit scales according to the M giant distance scale, and distances to
simulated debris particles similarly.  However, of course an orbit cannot
be scaled up or down in distance without affecting the velocities, and
thus the distance uncertainty necessarily implies an equivalent
velocity uncertainty.   

\begin{figure}[htbp]    
~\centering
\includegraphics[width=0.95\textwidth]{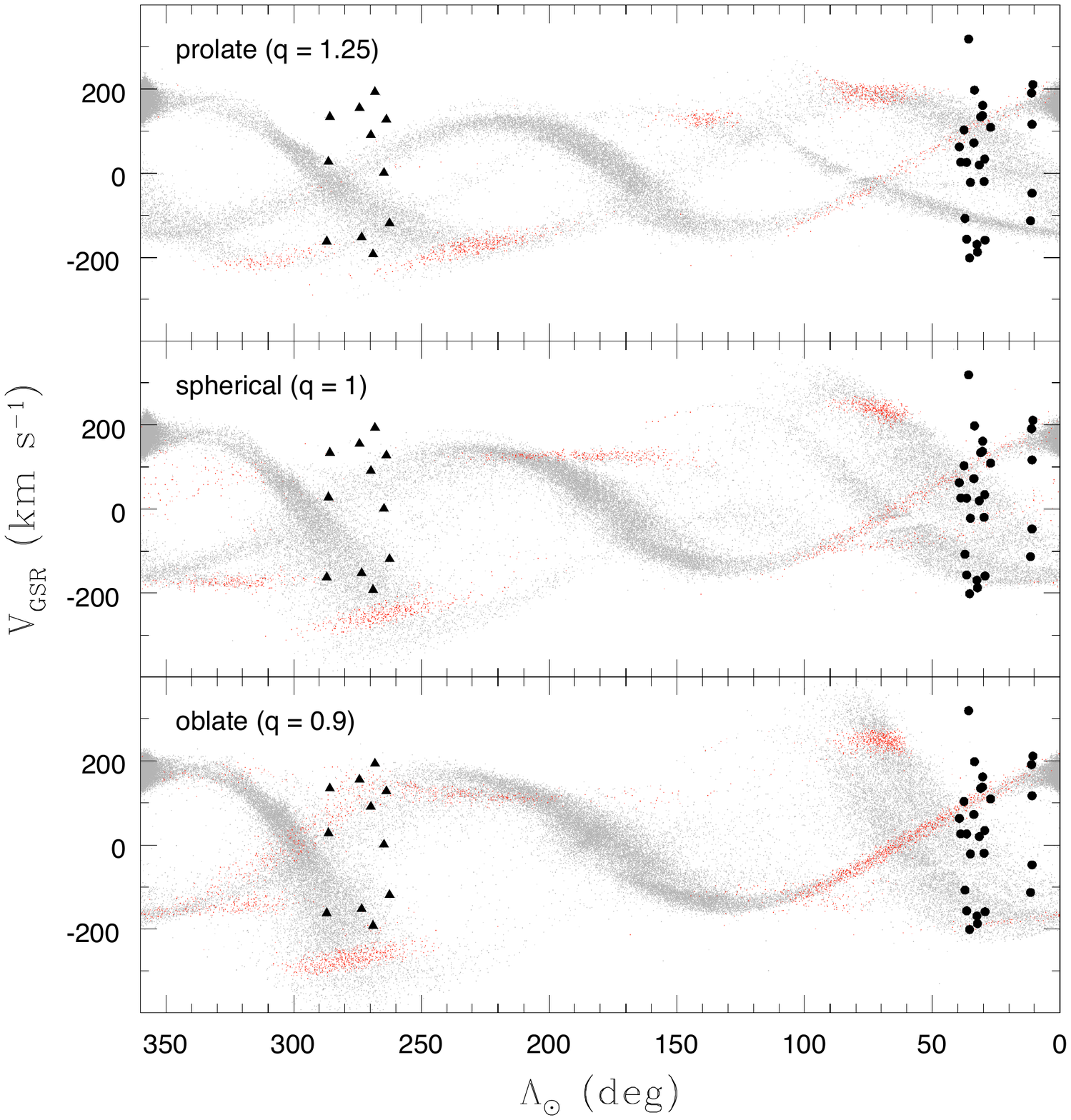}
        \caption[Radial velocities against $\Lambda_{\odot}$ from
          models with prolate, spherical and oblate Galactic halo
       potentials \citep{LJM05}, with values for hypothesized Sgr
       Stream members observed in the Sgr and VSS regions overplotted.
       Model distances 16--21 kpc are highlighted.]{Radial velocities
       against $\Lambda_{\odot}$ from models with prolate
       \textit{(top)}, spherical \textit{(middle)} and oblate
       \textit{(bottom)} Galactic halo potentials \citep{LJM05}.
       Red dots denote simulated
       stars with heliocentric distances $d$ = 16--21 kpc while gray
       denotes stars at all other distances.  Overplotted are values
       for the 26 observed RRLs in the 20 h and 21.5 h Sgr regions
       \textit{(circles)} and the 11 RRLs in the VSS region
       \textit{(triangles)}. All \Bsun\ values are shown. }
        \label{allmodelsd1621}        
\end{figure}

We also show, for comparison,
equivalent plots highlighting simulated particles at the observed
distances of 16--21 kpc in Fig.\ \ref{allmodelsd1621}. Again all
\Bsun\ values are included.
Overplotted on Figs \ref{allmodels} and \ref{allmodelsd1621} 
as filled circles are our
observational data in the Sgr region, with the 20 h region
corresponding to \Lsun\ $\approx$ 10\degree\ and the 21.5 h region
corresponding to \Lsun\ $\approx$ 25--40\degree.  In order to explore
their possible association with the Sgr Stream (as discussed in
\S\ref{intro_sgr}), our observed RRLs in the VSS region are also
plotted (triangles at \Lsun\ $\approx$ 260--290\degree).

\begin{figure}
~\centering
\includegraphics[width=0.95\textwidth]{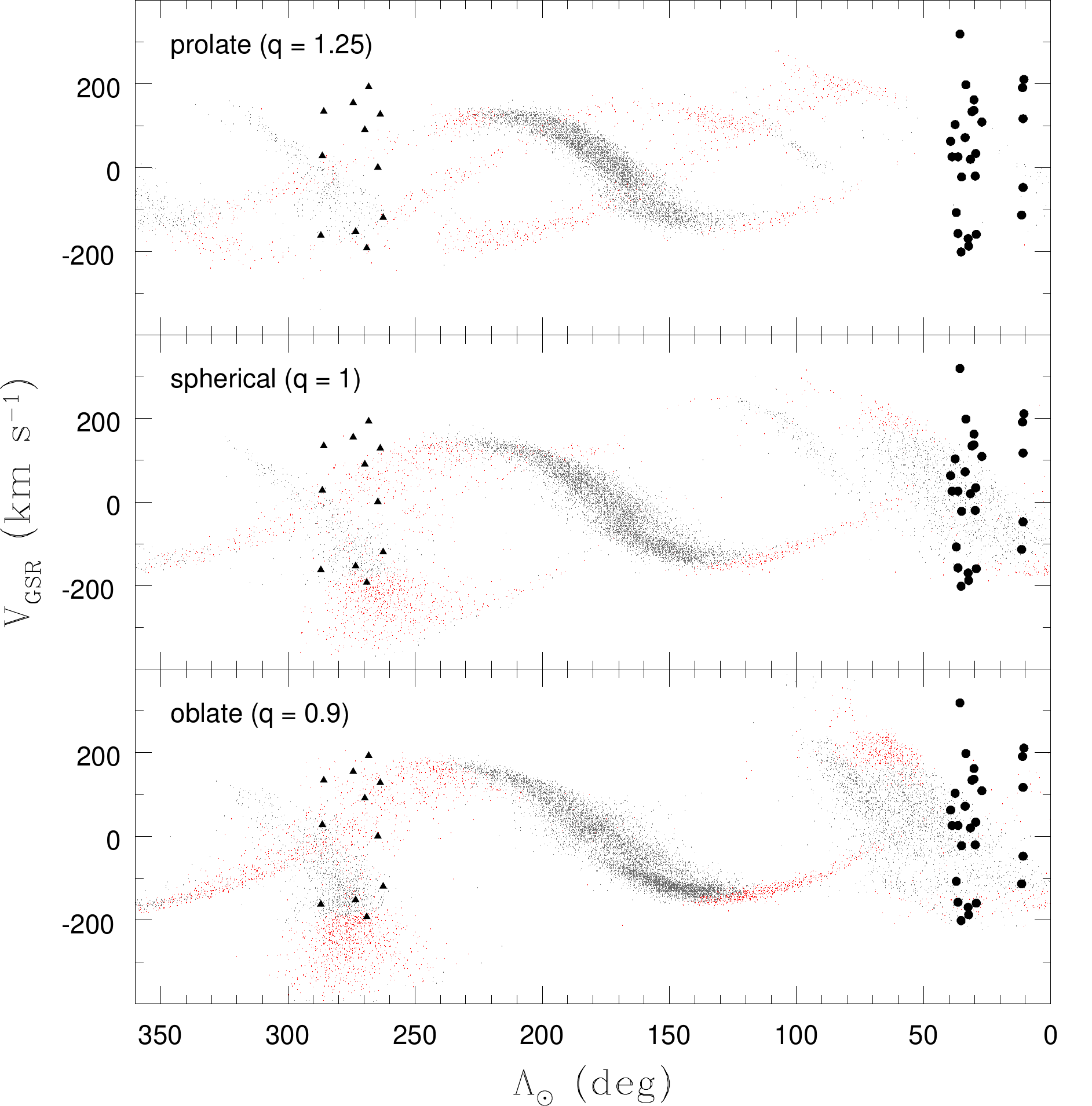}
\caption{As for Fig.\ \ref{allmodels}, i.e.\ model particle distances from 6--31 kpc, 
except that the model points plotted
have \Bsun\ values that fall within the SEKBO survey region. \label{new_fig6}}
\end{figure}

\begin{figure}
~\centering
\includegraphics[width=0.95\textwidth]{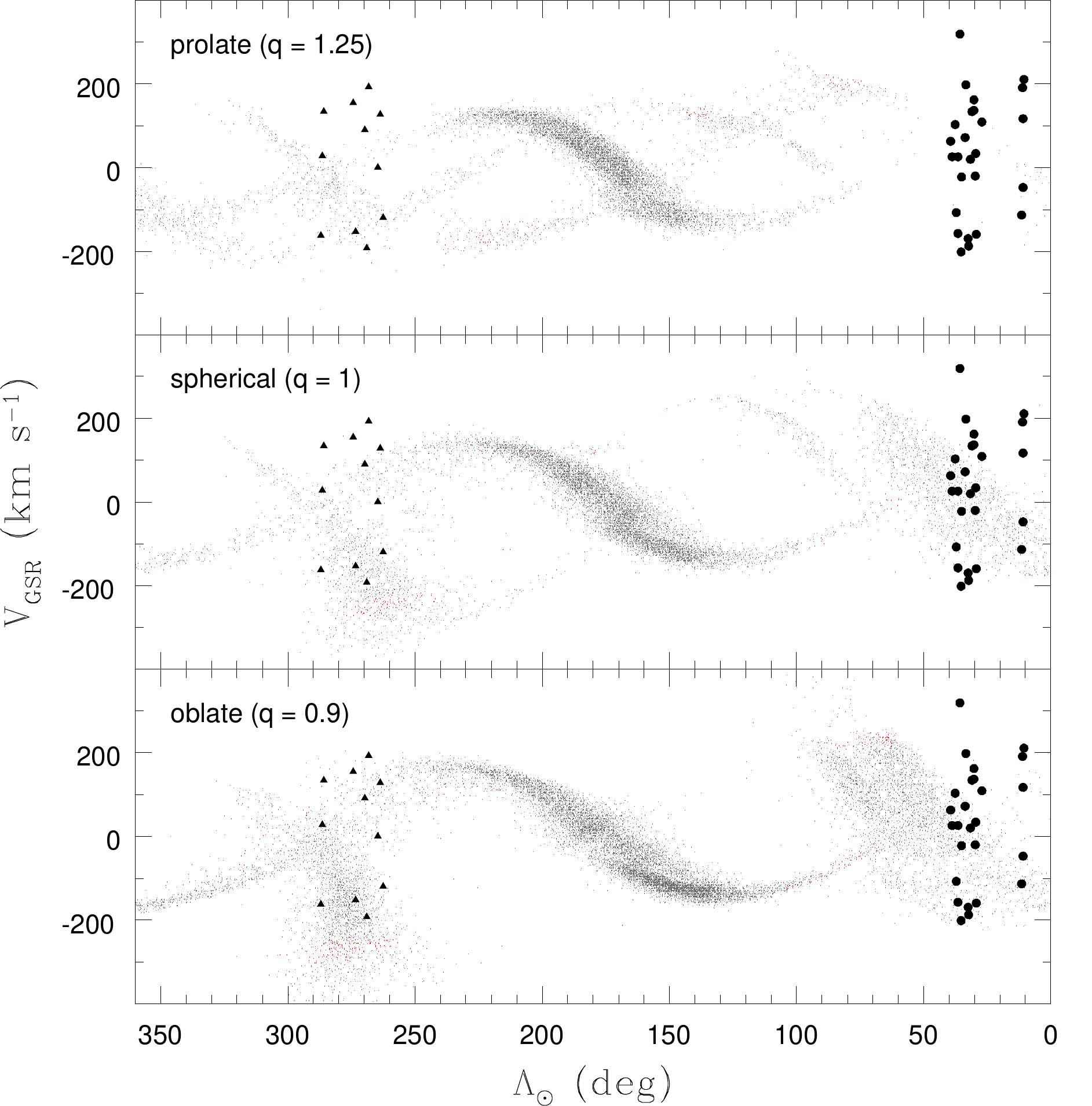}
\caption{As for Fig.\ \ref{allmodelsd1621}, i.e.\ model particle distances from 16--21 kpc,
except that the model points plotted
have \Bsun\ values that fall within the SEKBO survey region. \label{new_fig7}}
\end{figure}

In both Figs \ref{allmodels} and \ref{allmodelsd1621} the model particles
are plotted for all values of the latitude coordinate \Bsun.  However, it
is necessary to keep in mind that the RRLs observed here are drawn from the
SEKBO survey, which has the ecliptic as its mid-plane, not the Sgr orbit
plane.  Thus, the SEKBO RRLs do not sample all the \Bsun\ values
at any given \Lsun.  For example, the RRLs in the 20~h
and 21.5~h regions have --16.2 $\leq$ \Bsun\ $\leq$ --7.2 while those in the
VSS region  have 11.2 $\leq$ \Bsun\ $\leq$ 22.4 (cf.\ Table \ref{spectrosgr}).
Specifically, as discussed in KMP08, the SEKBO survey plane
cuts through the Sgr plane at an angle of $\sim$16$\arcdeg$ with an opening 
angle of 10$\arcdeg$.  Consequently, we show in Figs \ref{new_fig6} and 
\ref{new_fig7}
the same data as for Figs \ref{allmodels} and \ref{allmodelsd1621}, but now
with the model particles plotted only if they fall within a $\pm$5$\arcdeg$
window about the SEKBO survey mid-plane.  Nevertheless,
given that the definition of the \Bsun\ = 0 plane is model dependent, 
where appropriate we 
will continue to show figures with both the full \Bsun\ range and with the
restricted range required by the SEKBO survey characteristics.

\subsection{20 and 21.5 h Sgr regions}

A comparison between the observed and simulated data in the 20 and
21.5~h regions reveals several points of interest.  
Considering first the location of the five 20 h stars in 
Fig.\ \ref{allmodels}, it is tempting to associate the 3 stars with large
positive \vgsr\ values with the recent trailing debris of the models, the
location of which is largely independent of the halo potential adopted.
However, inspection of Fig.\ \ref{new_fig6} indicates that such an
interpretation is not clear-cut: the recent trailing debris in the models
lies relatively close to the \Bsun\ = 0 plane and thus the number
of model points from this feature is significantly reduced when the
restriction to the SEKBO survey region is applied, and   
for the case of the prolate halo model,
it is questionable as to whether the observed stars are related to Sgr at all.
These conclusions are not substantially altered with the more 
restricted in distance views shown in Figs \ref{allmodelsd1621} and 
\ref{new_fig7}.  We also note
that the distance to Sgr itself is $\sim$28 kpc \citep[e.g.][]{SL95} so that
the number of model points in the vicinity of \Lsun\ = 0 in 
Figs \ref{allmodelsd1621} and \ref{new_fig7} is naturally reduced 
compared to the figures with the wider distance range.

The larger number of stars in the 21.5 h sample, however, provides a stronger 
basis on which to make comparisons with the model predictions.  Adopting the
underlying smooth halo density distribution from Fig.\ \ref{new_fig1}, we 
expect that the 21.5~h sample has a field halo contamination of $\sim$40\%, or
8 $\pm$ 3 stars; the remainder are likely to be associated with Sgr.
In both Figs
\ref{allmodels} and \ref{allmodelsd1621} there is a group of
seven stars centered at \Lsun\ $\approx$ 35$^{\arcdeg}$ and \vgsr\ $\approx$
100 \kms\ whose variation of \vgsr\ with \Lsun\ suggests that they are
members of the recent trailing debris stream \textit{(debris-a)} regardless
of the halo potential adopted.  There are also, however, of order six stars
with --25 $\leq$ \vgsr\ $\leq$ 40 \kms\ which show no obvious connection
to any stream in any model.  Presumably a substantial fraction of these 
are field halo
objects, as is most likely the highest velocity star in the sample.  There 
is then an apparent gap
with no stars found between \vgsr\ $\approx$ --25 and \vgsr\ $\approx$ --110
\kms.  We suggested above that this gap may be statistically
significant, and based on the simulated particles it could plausibly be
interpreted as dividing members of the old leading debris stream
\textit{(debris-c)} from the old trailing stream \textit{(debris-b)}.
The most notable feature, however, is the
group of five stars with highly negative radial velocities (\meanvgsr =
$-175$ \kms, $\sigma$ = 19 \kms, slightly less than the uncertainty in
the measurements) which we noted as a distinct peak in Fig.\
\ref{genhisto_sgr}.  Under the smooth halo velocity distribution assumption 
stated above, we would expect a negligible number (0.4 $\pm$ 0.1) of the 
expected 5--11 contaminating halo field RRLs in the 21.5 h sample to have 
\vgsr\ values between --200 and --150 \kms.
The significance of the feature is thus beyond
question.  Comparing this observed group of RRLs with the simulated
particles in Figs \ref{allmodels} and \ref{allmodelsd1621}, it is clear 
that only the oblate
model predicts any significant number of stars with such velocities at
the appropriate distances.  In fact, the association of this group of stars 
with Sgr debris arguably rules out the prolate model for the halo
potential, at least in the context of the \citet{LJM05} models.

Once again, however, these possible interpretations need to be weighed 
when the observations and the models are considered in the context of the
SEKBO survey selection window (Figs \ref{new_fig6} and \ref{new_fig7}).
Identification of any of the observed stars with Sgr then becomes problematical
in the case of the prolate halo potential, as was noted above for the 20~h
sample.  Also as was noted for the 20~h sample, the models of LJM05 do not
predict any recent trailing debris at the \Bsun\ values of the 21.5~h sample
stars regardless of the adopted halo potential.  Significantly, however,
as seen in the bottom panel of Fig.\ \ref{new_fig6},
the LJM05 model with the oblate halo potential continues to predict the
existence of a population of stars with large negative \vgsr\ values at 
\Lsun\ values similar to those of the 21.5 hr sample, {\it even when the 
model points are limited to the SEKBO survey region}.  This remains the
case when the model points are more restricted in distance, as shown in
Fig.\ \ref{new_fig7}.  Indeed, in the context of the \citet{LJM05} models,
the identification of this group of stars
as Sgr debris strongly supports an oblate model for the halo potential.

\begin{figure}
~\centering
\includegraphics[width=0.75\textwidth]{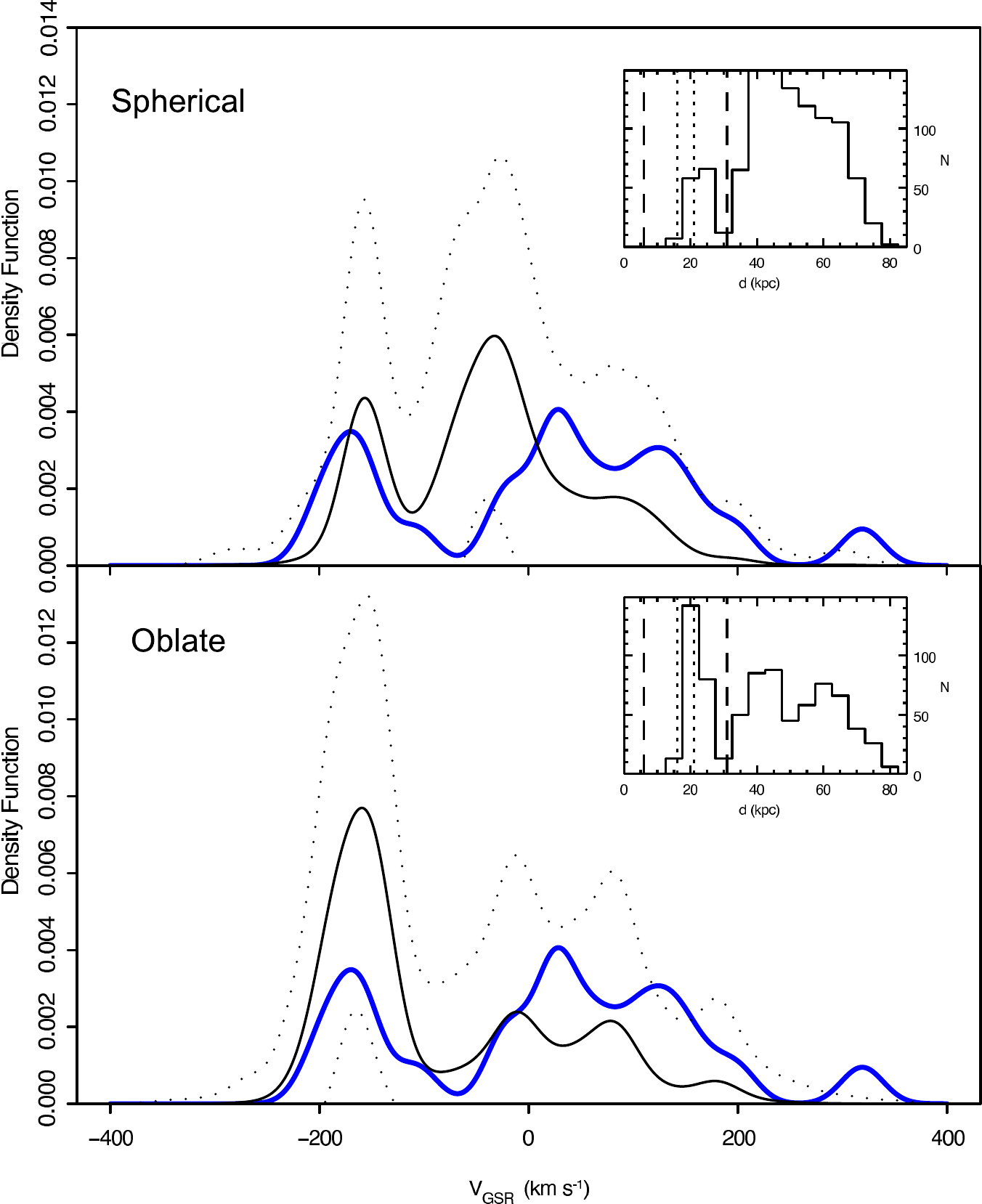}
\caption{Comparison of the generalised histogram for the observed
\vgsr\ velocities of the 21.5~h sample (blue solid line, cf.\ right panel of
Fig.\ \ref{genhisto_sgr}) with predictions from the LJM05 models.  The
upper panel is for the spherical halo potential model and the lower panel is
for the oblate model.  In each panel the black solid line is the mean 
generalized histogram from the ``model + field halo" trials (see text for 
details), while the dotted lines are the $\pm$3$\sigma$ bounds from the trials.
For both models the observed distribution lies within the trial boundaries.
The inset histograms show the distance distribution of simulated stars for 
each halo potential in the \Lsun\ and \Bsun\ range appropriate for the
observed sample.  Note that \citet{LJM05} added a 17\% artificial random 
distance scatter to the simulated particles to mimic the photometric
distance error in the 2MASS sample.  The vertical dashed lines indicate the distance 
range of the model particles used in the trials (6--31 kpc), while the vertical dotted
lines indicate the distance range of the observed stars (16--21 kpc).  \label{new_fig8} }
\end{figure}

To investigate the reliability of the qualitative interpretations 
presented above, we have conducted a number of Monte-Carlo trials.  In these 
trials we have chosen individual \vgsr\ values randomly from the set 
of model particle velocities, subject to: (1) the SEKBO selection window; 
(2) the \Lsun\ range of the 21.5 h observed sample; and (3) distance limits 
of 6-31 kpc (cf.\ Fig.\ \ref{new_fig6}).  We have also selected randomly from 
the assumed 
smooth halo velocity distribution, with the ``Sgr" and ``field halo" 
selections in the ratio 0.6 to 0.4.  The total selected sample for each trial
is set equal to that for the observed 21.5 h sample (i.e.\ 21 stars).  
The velocities are then convolved with the observed \vgsr\ errors of 
20 \kms\ and a generalized histogram constructed.  Multiple trials then
allow the generation of a mean generalized histogram and $\pm$3$\sigma$ limits
about that mean as a function of \vgsr.  We can then investigate the extent
to which the observed velocity distribution (cf.\ Fig.\ \ref{genhisto_sgr})
is consistent with the ``model + field halo" predictions.  The results are 
shown in Fig.\ \ref{new_fig8} for the LJM05 models with spherical and oblate 
potentials.  We cannot perform this test for the prolate models as there are 
simply insufficient model points meeting the criteria to allow adequate 
sampling.  Similarly, there are also insufficient model points to carry
out the simulations with the distances restricted to 16--21 kpc.

It is nevertheless apparent from Fig.\ \ref{new_fig8} that the observed data 
are largely consistent with the ``model + field halo" predictions for both the 
spherical and oblate halo potential models.  For the spherical model, the
largest difference between the observations and the model predictions lies
at \vgsr\ values of approximately --50 \kms, where the model predicts a
peak in the velocity distribution while the observations show a deficiency.
The discrepancy marginally exceeds the 3$\sigma$ bound but we do not 
regard it as strong enough to definitely rule out the spherical halo model.
In the oblate case, the biggest difference is that the model predicts a larger
number of stars at large negative \vgsr\ than are actually observed, although
the data are within the $3\sigma$ bound and there is close agreement 
as regards the location of the velocity peak.  Further, as distinct from the
spherical case, for the oblate model there is agreement with the observations
as regards the deficiency of stars around \vgsr\ $\approx$ --50 \kms.  As
noted above, it is tempting to identify this gap as separating the old leading
debris stream {\it (debris-c)} from the old trailing stream {\it (debris-b)}.

At this point it is also useful to consider the distance
distribution of the simulated particles in the different models.  The
inset distance histogram in Fig.\ \ref{new_fig8} shows that for the oblate 
model and the relevant \Lsun\ and \Bsun\ range, the density 
of the model particles peaks around 20 kpc and then falls off as the
distance approaches 30 kpc, with a second peak at distances $\sim$40
kpc.  The location of this peak in the model distribution is consistent
with the peak in the observed density distribution of the SEKBO RRLs seen
at this distance in Fig.\ \ref{new_fig1}. The spherical model, however, 
shows a less marked peak at $\sim$25 kpc with the majority of the model 
particles lying at $\sim$40 kpc and beyond.

In summary, the stars of the 21.5 h sample are most consistent with the 
predictions of the oblate halo potential model of LJM05.  In contrast, the 
prolate model does not predict any significant number of Sgr debris stars at 
the \Lsun\ and \Bsun\ values of the observed sample, and at the very least, 
requires modification if it is to be viable. The spherical potential model is
not inconsistent with the observations, but predicts stars with velocities
where the observations show a definite lack.  In all cases, however, the
observations seem to require that the recent trailing debris stream has a
larger spread in \Bsun\ than predicted by any of the models.
 
\subsection{12.4 and 14 h VSS regions}

Thus far we have only discussed RRLs in the 20 and 21.5 h Sgr regions.
We now turn our attention to the RRLs in the VSS region.
We note first that the SEKBO survey data \citep{KMP08}
indicates a significant overdensity at the distance and direction of these
RRLs, so the degree of contamination from field halo RRLs is likely to be
small.  The VSS RRLs from Paper~I are divisible into two groups: four stars 
with significant negative \vgsr\ values, and 7 stars with \vgsr\ values that 
range from near zero to $\sim$200 \kms.  For both groups it appears that the 
velocities are somewhat higher for lower \Lsun\ values.

To explore the potential association of these stars with Sgr debris streams
we once again consider their location in the 
context of the LJM05 models: Figs \ref{allmodels} and \ref{allmodelsd1621}
show the model data for all \Bsun\ values while Figs \ref{new_fig6} and
\ref{new_fig7} show the model data restricted to SEKBO survey range.
The VSS region RRLs are the 11 stars with \Lsun\ $\approx$ 260--290\degree. 

Turning first to the prolate case in Figs \ref{allmodels} and \ref{new_fig6},
we see that the 4 stars with significant negative \vgsr\ velocities lie in
the region of model particles from old leading debris {\it (debris-d)} 
(cf.\ Paper~I) and the velocity trend with \Lsun\ is consistent with that of 
the models.  This remains the case when the model \Bsun\ values are restricted
to the SEKBO survey region.   Are these stars, which have \meanvgsr\ =
--156 \kms, part of a real Sgr related substructure?  In this region, 
\citet{NYC07} detected a group of F stars
in SDSS data having a similar velocity (\vgsr\ $\approx$ --168 \kms).  More
recently, in QUEST data, \citet{VJZ08} found a significant number of
Blue Horizontal Branch stars at a similar velocity (\vgsr\ $\approx$ --171
\kms) as well as two RRLs.  They place the group at 11 $\pm$ 2 kpc,
but note that the stars are near the faint limit of their sample and
suggest that it may be the near side of a larger substructure.  They
believe it to be the same feature as that identified by
\citeauthor{NYC07}, which has a quoted distance of 11--14 kpc.
\citeauthor{VJZ08} note that the nominal 16\% distance uncertainty in
\citeauthor{NYC07}'s data may in fact be up to 40\%.  It is thus not
unreasonable to assume that the feature we have detected (at 16--21
kpc) is the same as that observed by both \citeauthor{NYC07} and
\citeauthor{VJZ08}  Based on their investigations, the latter
conclude that none of the halo substructures they detected were due to
the leading Sgr arm, nor did \citeauthor{NYC07} associate these stars
with Sgr debris.  However, we propose that this group of stars with
highly negative radial velocities is indeed associated with the
leading Sgr Stream, but that these stars were stripped on an
\textit{earlier} orbit (see \textit{debris-d}) than that considered by
\citeauthor{NYC07} and \citeauthor{VJZ08} \textit{(debris-c)}.  The velocities
and the velocity
trend favour the prolate model but we must keep in mind that the number of
observed stars is small.  On the other hand, in the prolate model the majority 
of the stars with positive \vgsr\ values, especially for the higher values, 
are not obviously related to Sgr debris. Thus in this situation, the VSS, 
which has \meanvgsr\ $\approx$ 130 \kms, would be unrelated to Sgr.  

For the spherical case, the identification of the negative \vgsr\ group with
old leading debris {\it (debris-c)} is mildly plausible though clearly the majority of 
model particles have more negative \vgsr\ values.  The lower \vgsr\ members of
the group of stars with
positive \vgsr\ values may be related to old trailing debris 
{\it (debris-b)} and the stars show the correct trend with \Lsun.  The highest 
velocity stars, however, are once again not obviously related to Sgr debris.  
These inferences remain unaffected when the restriction to the SEKBO survey 
is applied.  

Finally, for the oblate case, as for the
spherical model the negative \vgsr\ group might be identified as old leading
debris, though the majority of model points are at lower velocities.  However,
all the positive velocity stars fall among the broad swathe of old trailing
debris seen in this model at these \Lsun\ values, and the general trend of
higher \vgsr\ values with lower \Lsun\ values is broadly consistent with the
model predictions.   
In this situation it is possible that the VSS contains at least 
a component that is related to Sgr.  Again restricting the model
particles to lie within the SEKBO survey range (cf.\ Fig.\ \ref{new_fig6})
does not significantly modify these inferences.

If we now turn to Figs \ref{allmodelsd1621} and \ref{new_fig7} in which
the distance range of the model particles is restricted to correspond to 
that of the observed stars, some of the above inferences may need to be 
modified.  In particular, the identification of the negative \vgsr\ group
with old leading debris in the prolate model case is less evident, as the
model particles are generally at larger distances than the observed stars.
It is also evident that in the spherical and oblate cases there are
significant numbers of model particles at the appropriate distance at 
negative \vgsr\ values. However, the particle velocities are too low by 50--100 
\kms\ to agree with the observations even though the model and observed data
show similar trends with \Lsun.  
Whether there are modifications that could be made to the models to address
these situations is a question beyond the scope of the current paper.

One thing that is particularly noteworthy in the lower panels of 
Figs \ref{allmodelsd1621} and \ref{new_fig7}, i.e.\ the oblate models, though,
is that the \vgsr\ values for six of the seven RRLs with positive \vgsr\ 
values fall in with the old trailing debris model particles, and show the
same trend of increasing \vgsr\ with decreasing \Lsun\ as the model particles.
This is even true in Fig.\ \ref{new_fig7}, which has the highest restrictions
on the model particles: limited distance interval and within the SEKBO
survey range.  In Paper~I we identified four of these stars (those with
\vgsr\ values of 91, 128, 134 and 155 \kms; cf.\ Table 3 of Paper~I) with
the VSS, but it is evident from the lower panel of Fig.\ \ref{new_fig7} that
all six stars could be plausibly interpreted as Sgr old trailing debris
in the context of the oblate potential model of LJM05.  Again this suggests
that the VSS may well contain at least a component that is related to Sgr.

The results of \citet[][ hereafter DZV06]{DZV06} can be used to further 
investigate the situation.  These authors have measured radial velocities for 
18 RRLs in the VSS region.  Like our sample, their stars lie at distances 
between 16 and 20 kpc.  They have a somewhat smaller \Lsun\ range 
($258 \arcdeg \leq$ \Lsun\  $\leq 275 \arcdeg$, cf.\ Table 1) and have 
generally lower \Bsun\ values ($7 \arcdeg \leq$ \Bsun\ $\leq 16 \arcdeg$, 
median 12.5$\arcdeg$) compared to our VSS sample ($11 \arcdeg \leq$ \Bsun\ 
$\leq 22 \arcdeg$, median 17.7$\arcdeg$).
Given that the precision of the radial velocities are essentially identical,
the DZV06 data set is directly comparable with our own observations. There are 
two stars in common between the two datasets (see Paper~I), for which the
velocity measurements agree within the combined errors.  

The recent results of \citet{SH09} also provide further input.  In their 
pencil beam survey of the halo \citet{SH09} identified one group of 6 
metal-poor red giants
and 7 pairs of stars that have similar distances and velocities.
Two of the pairs, namely pair 7 and pair 8, have positions, velocities and 
distances that likely associate them with the spatial overdensity in
Virgo \citep{SH09}.  The stars have \Bsun\ values similar to those of the 
\citet{DZV06} RRLs.

\begin{figure}
~\centering
\includegraphics[width=0.95\textwidth]{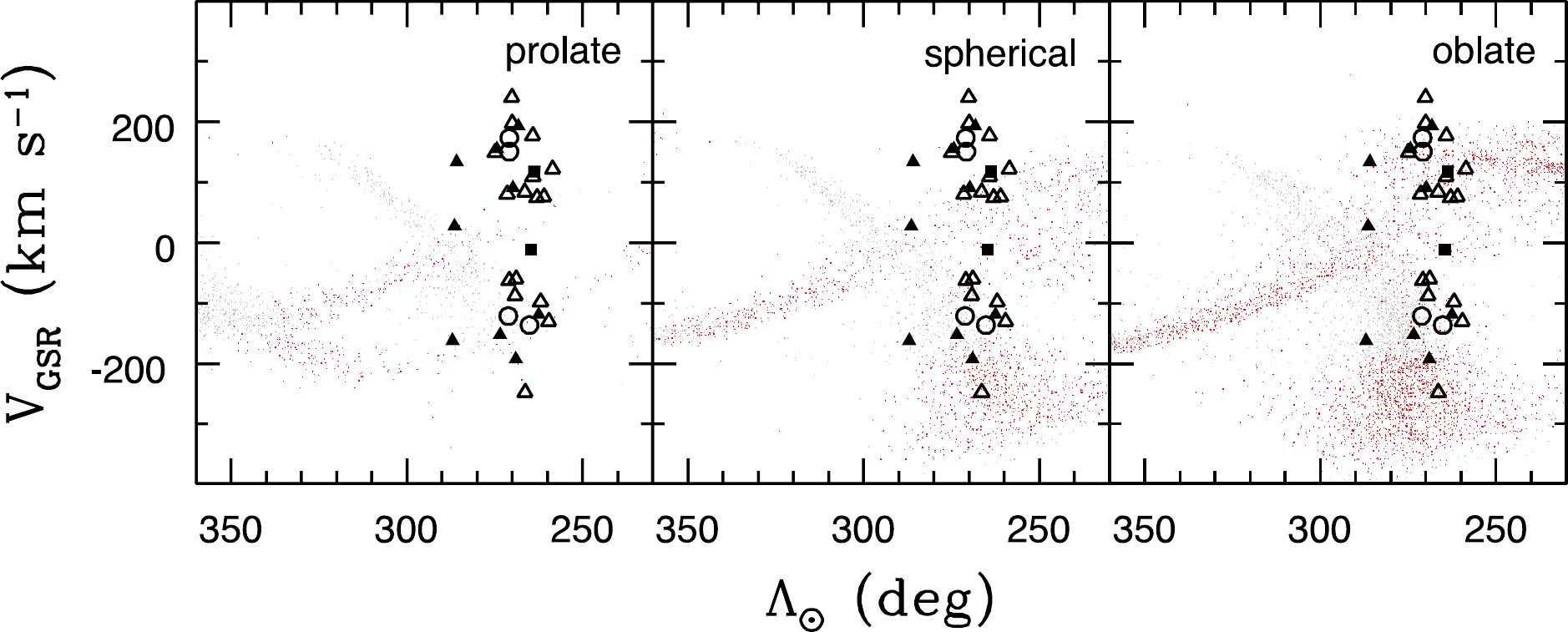}
        \caption{Radial velocities against
       $\Lambda_{\odot}$ from models with prolate \textit{(left)},
       spherical \textit{(middle)} and oblate \textit{(right)}
       Galactic halo potentials \citep{LJM05}.  Red dots
        denote simulated stars with
       heliocentric distances $d$ = 6--31 kpc while gray denotes particles
       at all other distances.  Only model particles with \Bsun\ values 
       exceeding 7$\arcdeg$ are shown.  Overplotted are values for our data
       (\textit{filled triangles}), for \citeauthor{DZV06}'s
       \citeyearpar{DZV06} RRLs in the VSS region (\textit{open triangles}), 
       and for \citeauthor{SH09}'s \citeyearpar{SH09}
       data for their pairs 7 and 8 (\textit{open circles}).  The two stars
       in common between our data and that of \citet{DZV06} are shown as
       filled squares at the average of the individual determinations.
       \label{new_fig9} }
\end{figure}

Fig.\ \ref{new_fig9} shows all three datasets compared to the LJM05 model data,
where we have plotted the model points for \Bsun\ values greater than 
7$\arcdeg$.  
Addition of the DZV06 and \citet{SH09} data generally supports the discussion 
above.  For the prolate case the stars from pair 7 of \citet{SH09} and two of 
the DZV06 stars with negative \vgsr\ values fall in the region
of model particles from old leading debris {\it (debris-d)}, and are
consistent with the trend of velocity with \Lsun\ of the model.  We note
again, however, that the model particles are larger distances than the 
observed stars (cf.\ Fig.\ \ref{allmodelsd1621}).  For the spherical case, only
the group of stars with \meanvgsr\ $\approx$ 100 \kms, i.e.\ those identified
with the VSS (DZV06, Paper~I), could readily be associated with Sgr, as
old trailing debris {\it (debris-b)}, and the trend of
\vgsr\ with \Lsun\ among the observed stars is once more consistent with the 
model prediction.  Again, however, the model particles are at larger distances
than the observed stars.  

For the oblate case, it is apparent that essentially
all the stars with positive \vgsr\ values could be interpreted as Sgr old
trailing debris.  The range of \vgsr\ values and the trend of these velocities 
with \Lsun\
are consistent with those of the model, and the association remains plausible
even when the distances of the model particles are restricted to those of the 
observed stars (see Figs \ref{allmodelsd1621} and \ref{new_fig7}).  
Indeed it is interesting to note that DZV06 also saw evidence of a possible 
radial velocity gradient in their VSS sample, when comparing their ``inner'' 
and ``outer'' samples (the outer sample has a lower mean \vgsr).

In Paper~I four of the RRLs shown in Fig.\ \ref{new_fig9} were identified as 
VSS members on the basis that their radial velocities were between 40 and 
160 \kms.  Similarly DZV06 label 8 of their 11 RRLs with positive \vgsr\ values
as probable VSS members, and pair 8 of \citet{SH09} is also readily identified
with the VSS\@.
There is a clear suggestion from Fig.\ \ref{new_fig9}, however, that the 
majority of the stars might in fact be part of a structure that varies 
considerably in radial velocity along its extent, with observed members having 
\vgsr\ as low as $\sim$20 \kms\ at \Lsun\ $\approx$ 295$\arcdeg$ and as high 
as $\sim$175 \kms at \Lsun\ $\approx$ 270$\arcdeg$.

In order to give these statements a more quantitative basis we have undertaken
Monte-Carlo simulations similar to those discussed above for the 21.5~h stars.
Specifically, we have investigated the extent to which the observed stars with 
positive \vgsr\ values could have resulted entirely from Sgr debris, at least
in the context of the \citet{LJM05} models.  We consider only the oblate model 
since, as is evident from Fig.\ \ref{new_fig9}, only this model has particles 
covering (almost) the full positive \vgsr\ range of the observed 
sample.  In the same way as for the 21.5~h sample trials, we have randomly 
chosen model particles
from the \citet{LJM05} oblate model subject to: (1) 258$\arcdeg$ $\leq$ \Lsun\
$\leq$ 275$\arcdeg$, which encompasses all but three of the observed sample; 
(2) \Bsun\ $\geq$ 7$\arcdeg$; and (3) distance limits of 6-31 kpc.  For each
selection the sample size is the same as that for the observed sample (28 stars)
and we have considered cases where the smooth halo background is either 
negligible, or makes up 25\% of the observed sample\footnote{DZV06 estimate the
contamination of their sample as $\sim$10--20 percent.}.  As before the smooth
halo velocity distribution is centered on \vgsr\ = 0 \kms\ with a dispersion
of 100 \kms.  A generalized histogram is then formed for the model velocities
with a kernel of 20 \kms\ as for the observations.  Multiple trials then
allow the generation of the mean generalized histogram predicted by the model
as well as $\pm$3$\sigma$ limits about the mean as a function of \vgsr.

\begin{figure}
~\centering
\includegraphics[width=0.95\textwidth]{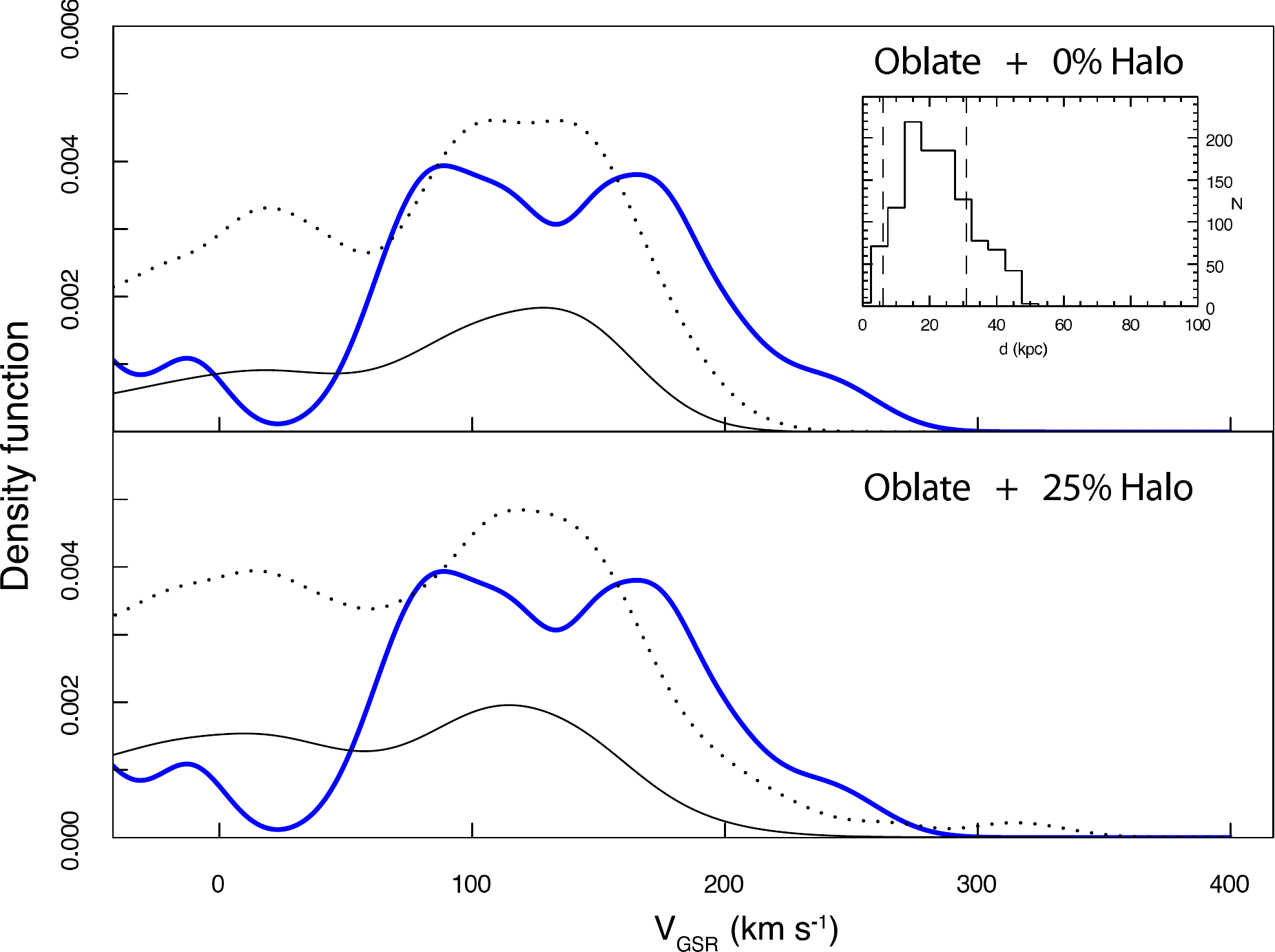}
\caption{Comparison of the generalized histogram for the sample of stars
in the VSS region (thick blue solid line) with the mean prediction of the LJM05 
oblate model (black solid line). The dotted lines are the $\pm$3$\sigma$
bounds from the trials.  The upper panel assumes no contribution
from a smooth halo background while the lower panel assumes a contribution
of 25 per cent.  The inset shows the distance distribution of the model
particles in the \Lsun\ and \Bsun\ range appropriate for the observed sample.
The dashed lines indicate the distance range of the model particles used
in the trials (6--31 kpc); the observed stars have distances between 13 and 
21 kpc.  It is evident that
the observed sample contains significantly more stars at positive \vgsr\
values than predicted by the model.
\label{new_fig10} }
\end{figure}

The results are shown in Fig.\ \ref{new_fig10} where the upper panel shows
the case of no smooth halo background while the lower panel assumes a 25\%
smooth halo contribution.  Given that it is already obvious from Fig.\
\ref{new_fig9} that oblate model predicts negative \vgsr\ values that are
considerably lower than those of the observed sample, we show the results
only for positive \vgsr\ values.
The insert in the upper panel shows the distance distribution of the model 
particles in the selected \Lsun\ and \Bsun\ range
with the dashed lines indicating the distance range of the model particles
used in the trials.  The observed stars have distances between 13 and 21 kpc
and this clearly corresponds to the peak in the distance distribution of
the model particles.  We also note that in the model there is essentially no difference 
between the distance distribution for the model particles with \vgsr\ $\leq$ 0
(i.e.\ old leading debris) and for the particles with \vgsr\ $\geq$ 0
(i.e.\ old trailing debris) indicating that in this region of the sky 
kinematic information is essential to separate the components.
The solid blue line is the velocity histogram
for the observed sample of 28 stars (6 stars from Paper~I, 16 stars from
\citet{DZV06}, 2 stars in common between Paper~I and DZV06 at their average
velocity, and the 4 stars from pairs 7 and 8 of \citet{SH09}).  The black
solid line is the mean prediction of the trials and the dotted lines are the
$\pm$3$\sigma$ bounds.

It is evident from Fig. \ref{new_fig10} that there is good correspondence
as regards distance between the model predictions and the observed sample,
and reasonable concurrence in velocity in that the model predicts a peak in
the distribution at \vgsr\ $\approx$ 120 \kms\ similar to what is observed.
However, the number of observed stars at positive \vgsr\ values is 
substantially 
larger than that predicted from the LJM05 oblate model, regardless of the 
assumed level of halo background contamination.  Indeed over the full
velocity range the trials
predict that there should be $\sim$3 times as many stars with \vgsr\ $\leq$ 0 
compared to stars with \vgsr\ $\geq$ 0 \kms\ (the factors are 3.4 for the no 
background
case and 2.6 for the 25\% smooth halo background case), whereas for the observed
sample, the factor is 0.75 (12/16).  In other words, based on the model, the
12 stars with \vgsr\ $\leq$ 0 \kms\ should correspond to only $\sim$4 stars
at positive \vgsr\ values, whereas 16 are observed.

This excess over the model predictions indicates that while Sgr debris from 
old trailing features may well contribute some stars with positive \vgsr\ 
values to the VSS region, at least in the context of the \citet{LJM05} models 
it cannot fully account for the observed excess.  Consequently, the VSS must 
contain a component independent of Sgr debris as has been argued by a number 
of authors \citep[e.g.][]{NYC07}.  

To summarize broadly the information from our VSS region stars and those of
DZV06 and \citet{SH09} as regards the shape of the halo, the stars with 
\vgsr\ values between approximately --100 \kms\ and
--200 \kms\ can be interpreted as old leading Sgr debris in the prolate halo
model of LJM05, though the model points lie at somewhat larger distances than 
the observed stars.  On the other hand, in the spherical and oblate models, the model
points agree better in distance with the observed stars, the model 
velocities are substantially more negative than those observed.
In contrast, the majority of the positive velocity stars agree 
in distance, velocity and in trend of velocity with \Lsun, with the predictions
of the oblate model for old trailing Sgr debris.  However, our simulations 
show
that the predicted number of stars with positive \vgsr\ values is substantially
less than the number observed.  Nevertheless, for these positive \vgsr\
stars, the prolate model 
is the least satisfactory representation of the data.
Thus, unlike the 21.5~h region where the support for the oblate model was
strong, the VSS region provides somewhat contradictory information regarding 
the flattening of the dark halo.

\subsection{Vivas et al.'s Sgr data}

\begin{figure}[t]    
~\centering
\includegraphics[width=0.95\textwidth]{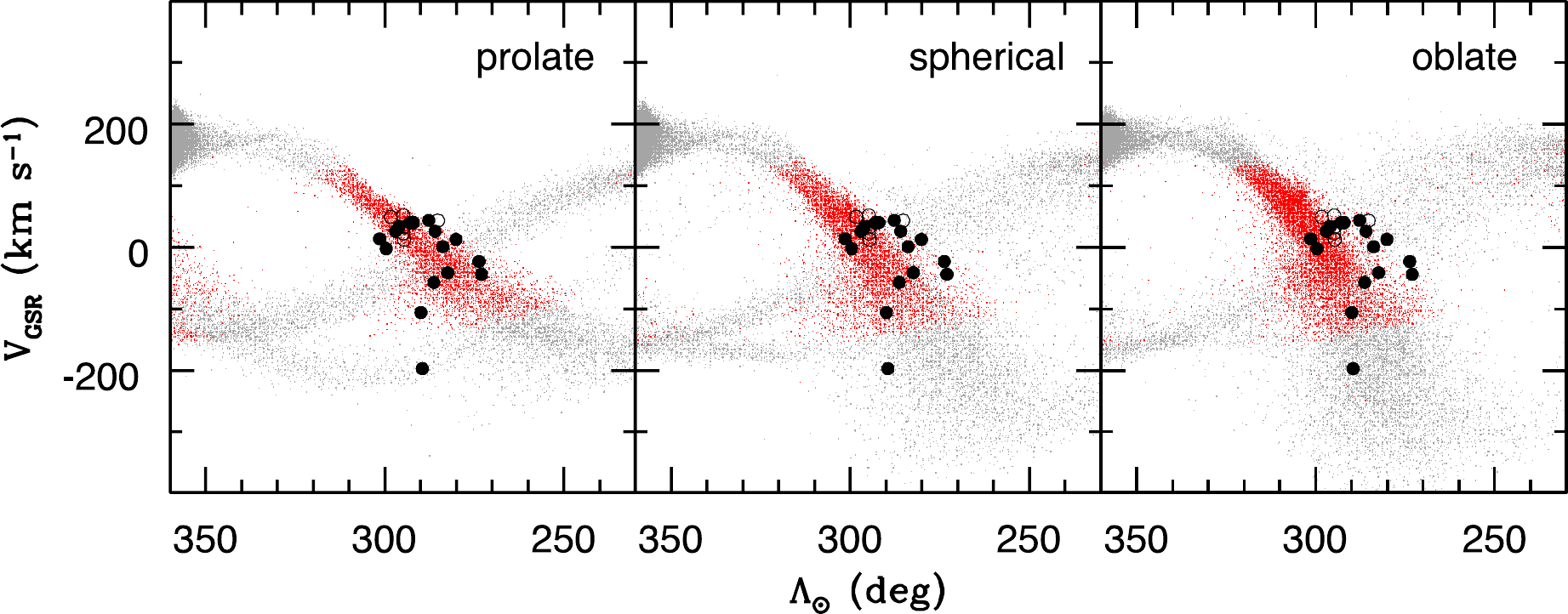}
        \caption[Radial velocities against \Lsun\ from
          models with prolate, spherical and oblate Galactic halo
       potentials \citep{LJM05}, with values for hypothesized Sgr
       Stream members observed by \citet{VZG05} overplotted.]{Radial
       velocities against \Lsun\ from models with prolate
       \textit{(left)}, spherical \textit{(middle)} and oblate
       \textit{(right)} Galactic halo potentials \citep{LJM05}.
       Red dots denote simulated
       stars with heliocentric distances $d$ = 40--65 kpc while gray
       denotes stars at all other distances.  All \Bsun\ values
       are shown.  Overplotted
       \textit{(filled circles)} are values for the 16 hypothesized Sgr
       Stream RRLs observed by \citet{VZG05}.  The \textit{open circles}
       represent the data for the Group 1 stars of \citet{SH09}, which
       are also likely associated with Sgr.}
        \label{allmodels_vzg}        
\end{figure}

As a further attempt to investigate the shape of the halo potential, we
compared the reported radial velocities of \citeauthor{VZG05}'s
\citeyearpar{VZG05} (hereafter VZG05) 16 RRLs to LJM05's models.
These RRLs are located at a different angular separation from Sgr to
our data, as shown in Fig.\ \ref{allmodels_vzg}.  The stars range
in distance from 45 to 62 kpc with a mean of 53 kpc so that in 
Fig.\ \ref{allmodels_vzg} we highlight in red model particles with distances
between 40 and 65 kpc.  The stars cover a \Bsun\ range from 9.6$\arcdeg$
to --8.3$\arcdeg$ so we are justified in plotting all \Bsun\ values for
the model particles.  We also note that the group of 6 probable Sgr stars 
identified by \citet{SH09} have locations, velocities and distances 
comparable to those of the VZG05 RRL stars.  Consequently, these stars are
also shown in Fig.\ \ref{allmodels_vzg}.  The \Bsun\ values for these stars
range between --11.5$\arcdeg$ and --6$\arcdeg$.

Upon comparing their data with those of \citet{Helmi04} and \citet{MGA04}, 
VZG05 concluded that spherical and prolate models fit better than oblate 
models, 
though none of the fits are completely satisfactory.  The comparison of 
LJM05's models with the VZG05 and \citet{SH09} data here reveals a similar 
conclusion.  
In Fig.\ \ref{allmodels_vzg}, the prolate model
is clearly the best fit, with all the stars except the two RRLs VZG05 
identified as probable non-members appearing to belong to the leading stream
\textit{(debris-c)}.  We note, however, that while the spherical and oblate
models produce progressively worse fits than the prolate model, they
are not completely inconsistent with the data.

\section{Metal Abundances} \label{metal_sgr}

\begin{figure}[htbp]    
~\centering
\includegraphics[width=0.8\textwidth]{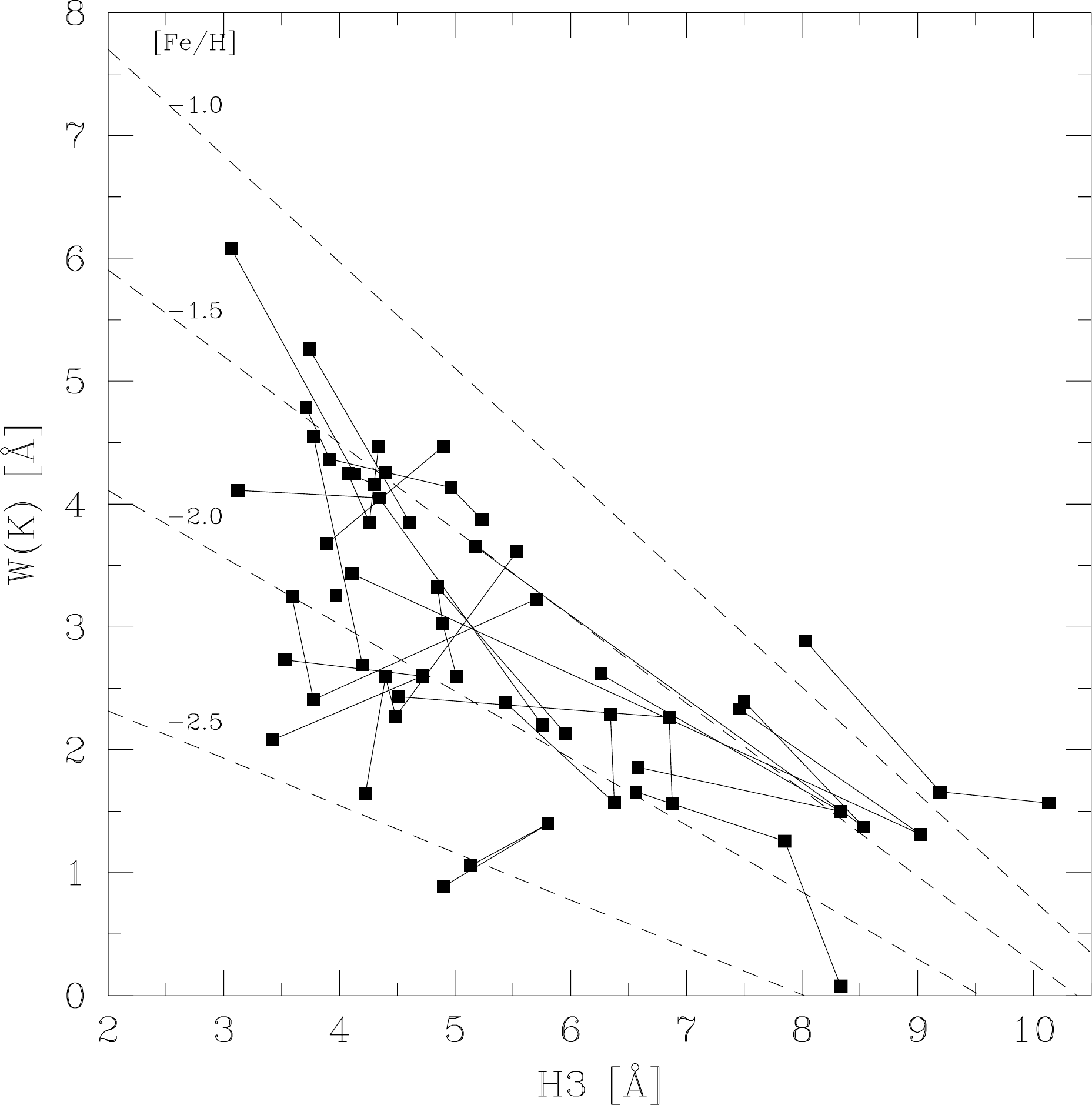}
        \caption[The pseudo-equivalent width of the Ca II K line
        against the average width of H$\delta$, H$\gamma$ and H$\beta$
        for the 21 type $ab$ RRLs in the Sgr region.]{The
        pseudo-equivalent width of the Ca II K line, corrected for
        interstellar absorption, against the average width of
        H$\delta$, H$\gamma$ and H$\beta$ for the 21 type $ab$ RRLs in
        the Sgr region.  Solid lines connect values for the same RRL
        observed at different phases.  The dashed lines are the loci
        of stars having the indicated [Fe/H] value according to
        \citeauthor{Layden94}'s \citeyearpar{Layden94} calibration.}
        \label{h3k_sgr}     
\end{figure}

As described fully in Paper I, metallicities ([Fe/H]) were calculated
for the type $ab$ RRLs using the \citet{FR75} method in which the
pseudo-equivalent width (EW) of the Ca II K line, $W$(K), is plotted
against the mean EW of the Balmer lines (H$\delta$, H$\gamma$ and
H$\beta$), H3.  RRLs trace out metallicity-dependent paths on this
plot as they vary in phase, as calibrated by \citet{Layden94}.  The
$W$(K)--H3 plot for the 21 type $ab$ RRLs in the 20 and 21.5 h Sgr
regions (note that one RR$ab$ star was omitted since all observations
corresponded to rising light phases) is shown in Fig.\ \ref{h3k_sgr}.

\begin{figure}[htbp]    
~\centering
\includegraphics[width=0.5\textwidth]{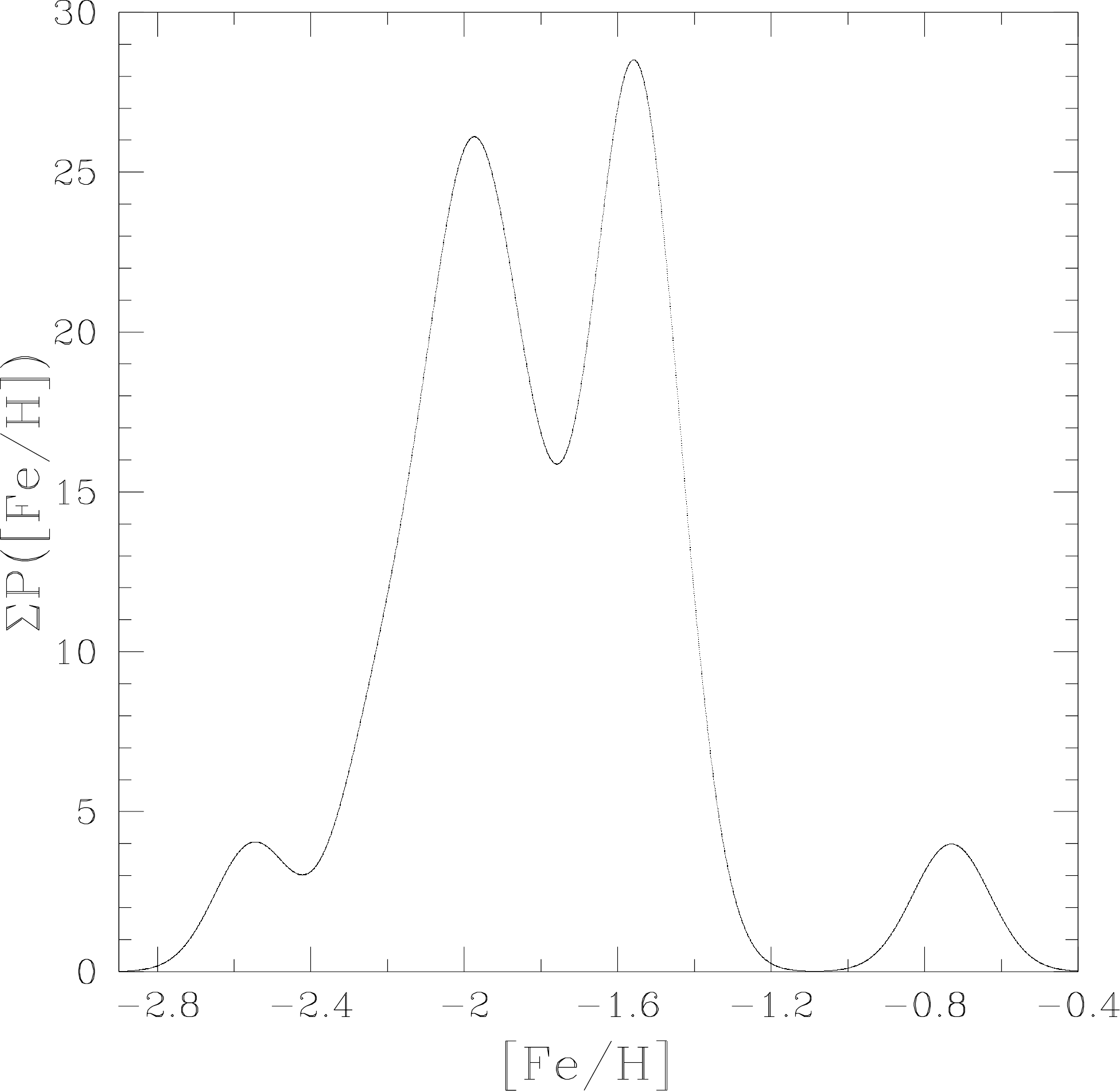}
        \caption[Generalized histogram of \feh\ for the 21 type $ab$
        RRLs in the Sgr region.]{Generalized histogram of [Fe/H] (with
        kernel of 0.1 dex) for the 21 type $ab$ RRLs in the Sgr region.}
        \label{fehgenhistosgr}        
\end{figure}

The values of [Fe/H] for these stars are listed in Table
\ref{spectrosgr} and their distribution is shown in Fig.\
\ref{fehgenhistosgr}.  Where more than one observation exists, the
tabulated values were calculated by averaging the [Fe/H] values from
the different phases (cf. Fig.\ \ref{h3k_sgr}).  Based on the stars
with multiple observations, the internal precision of a single [Fe/H]
determination is 0.11 dex.  For this sample, \meanfeh $= -1.79 \pm
0.08$ dex on our [Fe/H] system with a dispersion of $\sigma$ = 0.38
dex (see Fig.\ \ref{fehgenhistosgr}).  This mean value is in close
agreement with VZG05's measurement of \meanfeh = $-1.76$, $\sigma$ =
0.22 dex, for 14 QUEST RRLs in the Sgr tidal stream.  As noted by
VZG05, the observed \meanfeh\ is consistent with the age-metallicity
relation of the main body of Sgr \citep{LS00} if these RRLs are coeval
with the oldest stellar population in the body, as they are indeed
expected to be.

We noted above that the 6 stars in the Sgr group (Group 1) of \citet{SH09} 
lie in the same part of the Sgr stream as the VZG05 RRL stars. These red giants
have a mean abundance of \meanfeh\ = --1.68 $\pm$ 0.15, in excellent accord 
with that for the VZG05 stars.  The observed abundance dispersion is 0.38 dex,
which, given the large ($\sim$0.3 dex) abundance uncertainties, implies an
intrinsic abundance disperion of $\sim$0.24 dex.  This is again in good accord
with the VZG05 results.  The most metal-poor star in the group has
[Fe/H] = --2.33 $\pm$ 0.31 while the most metal-rich has [Fe/H] = --1.29 
$\pm$ 0.26 dex.

The \feh\ dispersion of our sample is larger than that found by VZG05 and
that of the \citet{SH09} group 1 stars,
and is closer to the dispersion of $\sigma$ = 0.4 dex found by
\citet{KCC00} for RRLs in the halo.  It should be recalled, however,
that our RRL sample likely includes field RRLs which are not part of
the Sgr Stream.  Contamination by such RRLs could thus inflate our
estimate of the \feh\ dispersion for Sgr Stream stars.  The dispersion
could also be disproportionately influenced by outliers.  We note that
if the most metal rich and the most metal poor RRL (visible as the
outermost, small bumps on Fig.\ \ref{fehgenhistosgr}) are excluded,
the remaining stars have $\sigma$ = 0.23 dex and an abundance range of
0.8 dex, consistent with the values of VZG05 and for the \citet{SH09} group
1 stars.  Including those two
stars, on the other hand, yields a much larger abundance range of 1.8
dex.  A more rigorous way of reducing the effect of outlying, extreme
values on the dispersion is to consider the inter-quartile range
(IQR).  We obtain IQR = 0.46 dex which is larger than that for VZG05's
data, IQR = 0.22 dex.  It thus appears that the larger \feh\
dispersion in our Sgr Stream sample than in VZG05's is not driven
solely by outlying values.
The differing abundance dispersions are perhaps unsurprising given
that we are sampling a different part of the Sgr Stream than VZG05 and the
group 1 stars of \citet{SH09}.  Nevertheless the close agreement of the
mean abundances for the three samples is intriguing. 

Fig.\ \ref{fehgenhistosgr} appears to show a hint of bimodal structure.
However, a KS-test shows that the null hypothesis, namely that the sample 
(excluding the two outliers) is drawn from a single gaussian distribution with 
the sample mean and standard distribution, cannot be excluded with any
significance: the probability that the null hypothesis is true is at least 
20\%.  Similarly, if we isolate the stars that contribute to the two apparent
peaks at [Fe/H] $\approx$ --2.0 and [Fe/H] $\approx$ --1.55, we find that
the four 20~h stars are split between the two peaks 3 stars to 1, and the fifteen 
21.5~h stars are split 8 stars  to 7, i.e.\ the apparent abundance separation does 
not show any correlation with location.  Further, for the 21.5 h stars, there is no 
difference in mean velocity or velocity dispersion between the groups 
separated by abundance.
  
The fact that the mean abundance we find for the 20 and 21.5~h RRLs
agrees with that from VZG05 (and that from the \citet{SH09} group 1 stars)
in a different part of the Sgr stream raises the question of whether
there is any pattern in mean \feh\ according to the part of the stream
to which the RRLs belong.  Indications of an age/metallicity gradient
along the Sgr Stream have been observed, with stars stripped from Sgr
on past perigalactic passages being older and more metal poor than the
Sgr core and stars stripped more recently
\citep{MSW03,MGA04,BNC06,CM07}.  

We thus searched for evidence of this
abundance gradient by comparing the metallicities of RRLs in the 20
and 21.5 h regions which appear associated with the recent trailing
debris stream (\textit{debris-a}) to those likely associated with the
older trailing debris stream (\textit{debris-b}).  The six RRLs in the
former have \meanfeh\ = $-1.76$, $\sigma = 0.24$ dex while the seven
stars in the latter have \meanfeh\ = $-1.68$, $\sigma = 0.48$ dex.
Thus, we find no evidence for a correlation between location on the
stream (and hence age of stripping) with abundance in our RRL sample.

A comparison of the recent results of \citet{WEB09} with those above
might suggest a different conclusion.  \citet{WEB09}
listed a mean abundance of \meanfeh\ = $-1.41 \pm
0.19$ for a sample of likely Sgr RRLs identified from their analysis of the
SDSS Stripe 82 region.  In the context of the LJM05 models their RRL sample
has a strong component from
recent trailing debris (\textit{debris-a}), especially if the halo potential
is oblate.  The majority of the abundance determinations in their sample come
from analysis of the periods and light curve shapes derived from the 
photometry \citep{WEB09}.  As such they are 
relatively more uncertain ($\sigma$([Fe/H]) $\approx$ 0.25 dex for the type ab, 
and 0.38 dex for 
the type c RRLs, \citet{WEB09}) than the direct spectroscopic determinations 
discussed here.  Thus any comparison of the \citet{WEB09} mean abundance with 
those for the other RRL samples discussed here should be treated with 
caution.  Nevertheless, given 
the association of the \citet{WEB09} sample with recent trailing debris, its
higher mean metallicity compared to those of the other Sgr RRL samples
might indicate the presence of an abundance gradient, not withstanding the
results discussed in the previous paragraph.
However, it must be recalled that RRLs are members of an old
population and thus RRL samples do not necessarily represent an
unbiased selection of stream members. 

\section{Discussion}

The results presented in this paper impinge on two related issues.  The
first is the extent to which the velocities and distances of
observed Sgr debris stars can be used in conjunction with the models of
LJM05 to place constraints on the shape of the Galaxy's dark halo.  The
second revolves around the interpretation of the overdensity in Virgo
and its potential relation to Sgr debris.

The available observational data and the analysis presented here indicate
that the Galactic halo substructure in the direction of Virgo is evidently
quite complex.  In Paper~I we showed that considered as a spatial overdensity,
the feature is large and diffuse and extends well to the south of the
declination limits of the SDSS and of the QUEST survey \citep[see also][]{KP09}.
However, kinematically, there appear to be at least two distinct components
that overlap spatially.  One is defined by the group of RRLs with \meanvgsr\
$\approx$ --160 \kms, which we identify as likely resulting from Sgr old
leading debris \citep[cf.][]{MPJ07}.  This kinematic signature is also seen 
in the work of \citet{NYC07}, \citet{VJZ08} and pair 7 of \citet{SH09}. 

At positive \vgsr\ values, there is concurrence in both spatial location, 
velocity, and velocity trend with orbital longitude of the combined 
observational sample of Paper~I, \citet{DZV06} and \citet{SH09} pair 8 with 
the predictions for Sgr old trailing debris in the LJM05 oblate model 
(cf.\ Fig.\ \ref{new_fig9}).  This is suggestive that such debris may play a
role in the (kinematically defined)
feature labeled by \citet{DZV06} as the Virgo Stellar Stream. 
Such a possibility was foreshadowed by \citet{MPJ07}.  However, our
simulations based on the LJM05 oblate model suggest strongly that Sgr old
trailing debris cannot be the sole source of the positive \vgsr\ stars.  
Instead it is likely that there is at least one other independent
structure that conspires with the Sgr debris to produce the observed excess.
Clearly a detailed spatial and kinematic survey of a large region of sky is
needed to clarify the situation.  The SEKBO survey catalog alone provides a 
more than ample selection of RRL candidates in the VSS region for follow-up 
(cf.\ Paper~I).  We also have underway a survey targeting red giants in
this region, which takes advantage of the wide field and large multiplex factor 
of the AAOmega multi-fiber spectrograph at Anglo-Australian Telescope.

As regards the shape of the halo potential, the existence of a distinct
group of RRLs with large negative \vgsr\ values in the 21.5~h sample strongly
favors the oblate model of LJM05.  However, the negative \vgsr\ stars in the
VSS region, interpreted as Sgr old leading debris, and the more distant 
samples of \citet{VZG05} and \citet{SH09} group 1, tell a different story in 
which prolate models are favored.  These findings echo those of LJM05,
wherein trailing data favor oblate models and leading data favor
prolate models, and reinforce the conclusion that the issue of dark
halo shape cannot be definitively resolved with current models of the
Sgr disruption.

LJM05 suggest that an evolution of the orbital parameters of Sgr over
several Gyr may need to be considered.  In addition, the Sgr dwarf
itself may need to be modeled as a two component system, in which the
dark matter is bound more loosely than the baryons.
\citeauthor{WKP08}'s \citeyearpar{WKP08} N-body simulations of
satellite disruption within ``live'' (cosmological) host halos suggest
that the situation may be far more complicated than previously
envisioned.  Their preliminary studies find little correlation between
debris properties and host halo properties such as shape.  They note
that the host dark halo undergoes a complex mass accretion history and
also comment that it cannot be easily modeled as a simple ellipsoid
due to the wealth of substructure present.  In contrast, \citet{SV08}
investigate the effects of dark matter substructure on tidal streams
and find that halo shape and orbital path play a much more important
role in the large-scale structure of the debris.  However, they do
note that substructure increases the clumpiness of the debris and
changes the location of certain sections compared to the predictions
from a smooth halo model.  This could add an extra complication to
studies such as the current one, where attempts are made to compare tidal
debris models with data.  

\section{Conclusions}

Analysis of follow-up spectroscopy of 26 photometrically confirmed
RRLs from a candidate list based on SEKBO survey data reveals a radial
velocity distribution which does not appear consistent with a smooth
halo population.  Based on their location, the RRLs are likely to be
associated with debris from the disrupting Sgr dwarf galaxy.  The 21
type $ab$ RRLs in the 20 and 21.5 h regions have \meanfeh\ = $-1.79
\pm 0.08$ on our system and a large abundance range of $\sim$1.8 dex
($\sigma$ = 0.35), or 0.8 dex ($\sigma$ = 0.23) omitting the two most
extreme values.  The interquartile range is 0.46 dex.  While the abundance
spread in our data appears to be larger, the mean
metallicity is consistent with that of \citet{VZG05} for Sgr tidal
debris RRLs, and that of the group 1 (Sgr) red giants from \citet{SH09},
which lie in a different part of the Sgr debris stream.
The mean abundance for the Sgr RRLs in the \cite{WEB09} sample, however,
is apparently somewhat higher, though based on a different technique. 

A comparison of the radial velocities with those predicted by the
models of \citet{LJM05} supports the hypothesis that the observed
RRLs predominantly belong to the Sgr Stream.  In the 21.5 h region, a
group of stars with highly negative radial velocities (\vgsr $\sim
-175$ \kms) is consistent with predictions for old trailing debris
when the Galactic halo potential is modeled as oblate.  In contrast, the
prolate model does not predict any significant Sgr debris at the \Lsun\
and \Bsun\ of the observed sample, indicating that it requires modification
if it is to be viable.  The observations also seem to require that the
recent trailing debris stream has a larger spread in \Bsun\ than predicted
by any of the models.

Comparison of radial velocities of VSS region RRLs with the Sgr debris
models reveals intriguing similarities in trends with \Lsun.  Together
with the evidence of spatial coincidence of these stars with the
predicted debris streams, our results provide observational support
for \citeauthor{MPJ07}'s \citeyearpar{MPJ07} proposition that Sgr
debris is in fact at least partially responsible for the overdensity
in Virgo.  In particular, it seems likely that the stars with large 
negative \vgsr\ values are best interpreted as old leading debris.
However, at least in the context of the LJM05 models, it appears unlikely
that the feature at \vgsr\ $\approx$ 100 \kms, the VSS, can be explained
solely as Sgr old trailing debris.  While Sgr old trailing debris may
make a contribution, the (kinematically defined) VSS is apparently a 
structure in the halo of the Galaxy independent of Sgr debris.  

Considering all the data sets for suspected Sgr Stream
members presented in this paper, we find further evidence for
\citeauthor{LJM05}'s observation that trailing debris is best fit by
oblate models while leading debris favors prolate models.  That is, we
are in agreement with \citeauthor{LJM05}'s conclusion that no single
orbit and/or potential can fit all the observed data.  Further
modeling is needed to investigate higher order effects such as orbital
evolution.  To this end, data from RRLs may prove useful as these old
stars could potentially have been stripped from Sgr several orbits
ago.  It may also be necessary for models to include the possibility that the
flattening of the halo varies with radius \citep[e.g.][]{BS05} or that it is triaxial
\citep[cf.][]{LM09}.  The high accuracy of determined distances to RRLs provides an
extra incentive to use these stars as probes of Sgr debris.

\acknowledgments

This research has been supported in part by the Australian Research
Council through Discovery Project Grants DP 0343962 and DP 0878137. We
thank the anonymous referee for comments on the original manuscript that led 
to a number of improvements in the presentation.  We are also grateful to
Prof\ Bob Zinn for supplying details of the \citet{DZV06} RRL stars and to
LJM05 for making the details of their model data publicly available.



{\it Facilities:} \facility{SSO:1m (WFI)}, \facility{SSO:2.3m (DBS-B)}





\end{document}